\documentclass[12pt,a4paper]{article}
\usepackage{color,graphics,amsmath,epsfig,rotating}

\def\VERSION{3.X}

\begin{document}

\newcommand{\ds}{\texttt{DarkSusy}}
\newcommand{\micro}{\texttt{micrOMEGAs}}
\newcommand{\py}{\texttt{PYTHIA}}
\newcommand{\HE}{\texttt{HERWIG}}
\newcommand{\msun}{{\rm M_{\odot}}}
\newcommand{\ie}{{\it i.e.} }
\newcommand{\eg}{{\it e.g.} }
\newcommand{\pbar}{${\rm \bar{p}}$}
\newcommand{\dbar}{${\rm \bar{D}}$}
\newcommand{\beq}{\begin{equation}}
\newcommand{\eeq}{\end{equation}}
\def\rsun{r_\odot}
\newcommand{\noi}{\noindent}

\newcommand{\calchep}{\texttt{CalcHEP}}
 \def\lnTop{\log\left(\frac{m_t^2}{Q^2}\right)}
\def\lsp{\chi}
\def\mlsp{m_{\chi}}
\def\mglu{m_{\tilde g}}
\long\def\symbolfootnote[#1]#2{\begingroup%
\def\thefootnote{\fnsymbol{footnote}}\footnote[#1]{#2}\endgroup} 

\begin{center}

{\Large\bf micrOMEGAs$\_ 3$ : a program for calculating dark matter observables} \\[8mm]

{\large   G.~B\'elanger$^1$, F.~Boudjema$^1$, 
A.~Pukhov$^2$, A.~Semenov$^3$ }\\[4mm]

{\it 1) LAPTH, Univ. de Savoie, CNRS, B.P.110,  F-74941 Annecy-le-Vieux, France\\
    2) Skobeltsyn Inst. of Nuclear Physics, Moscow State Univ., Moscow 119992, Russia\\
      3) Joint Institute of Nuclear research, JINR, 141980 Dubna, Russia  }\\[4mm]

\end{center}

\begin{abstract}
\micro~ is a code to compute dark matter observables  in generic extensions of the standard model.
This new version of \micro~ is a major update which includes a generalization of the Boltzmann equations to accommodate models with asymmetric dark matter or with  semi-annihilation and a first approach to a generalization of the thermodynamics of the Universe in the relic density computation. Furthermore a switch  to include virtual vector bosons in the final states  in the annihilation cross sections or relic density computations is added. Effective operators to describe loop-induced couplings of Higgses to two-photons or two-gluons are introduced and reduced couplings of the Higgs are provided allowing for a direct comparison with recent LHC results. 
A  module that computes the signature  of DM captured in celestial bodies in neutrino telescopes is also provided. 
Moreover the direct detection module has been improved as concerns the implementation of the strange ``content" of the nucleon. New extensions of the standard model are included in the distribution. 
\end{abstract}

\section{Introduction}

Cosmological and astrophysical observations have established that the matter content of the Universe is dominated by dark matter (DM).  However the nature of this dark matter still needs to be elucidated, therefore searches for dark matter signals are pursued actively.
Dark matter observables include the relic density,  the elastic scattering cross section on nuclei, the indirect signatures of DM annihilation in the galaxy (leading to photons and antimatter) or DM capture in celestial bodies (leading to neutrinos).

For all these observables, experiments are currently taking data and improved results are expected. 
The PLANCK satellite has recently determined the DM relic density with a precision of 2\%~\cite{Ade:2013lta}.
Direct detection experiments such as DAMA~\cite{Bernabei:2010mq}, CDMS~\cite{Agnese:2013rvf} CoGeNT~\cite{Aalseth:2011wp} and CRESST~\cite{Angloher:2011uu} have found events that could be compatible with DM in the range 5-30 GeV, these are however incompatible with the 
upper limits on spin  independent DM interactions with nuclei set by   Xenon100~\cite{Aprile:2012nq}. Furthermore COUPP~\cite{Behnke:2008zza},  Picasso~\cite{Archambault:2009sm} and Xenon100~\cite{Aprile:2012nq} have set limits on  spin dependent interactions. Larger direct detection projects with ton scale detectors are planned, e.g. Xenon,  and should improve by more than one order of magnitude the current sensitivity.  
Indirect detection experiments are  measuring  the spectrum of cosmic rays in different channels. 
 FermiLAT has measured the photon flux from the Galactic Center and from Dwarf Spheroidal Galaxies, the latter have allowed to probe the canonical DM annihilation cross section for DM lighter than 40GeV~\cite{Ackermann:2011wa}. Furthermore an anomaly corresponding to a gamma-ray line at 130 GeV  has been extracted from the FermiLAT data~\cite{Bringmann:2012vr,Weniger:2012tx}. FermiLAT~\cite{FermiLAT:2011ab}, PAMELA~\cite{Adriani:2008zr} and AMS~\cite{Aguilar:2013qda} have measured the positrons fluxes, showing an anomaly at high energies while PAMELA data on anti-protons are in perfect agreement with expectations from the background~\cite{Adriani:2008zq}. AMS  will pursue the measurement of antimatter (positrons, antiprotons and antideuterons) and photon fluxes. Despite some anomalies, none of these searches lead to conclusive evidence for DM.
  Finally neutrino signals from annihilation of dark matter captured by the Sun are being probed by neutrino telescopes such as SuperK~\cite{Tanaka:2011uf}, ANTARES~\cite{Zornoza:2012xn} or ICECUBE~\cite{IceCube:2011aj}. 
 
On a different front, the LHC has discovered the Higgs boson~\cite{Aad:2012tfa,Chatrchyan:2012ufa} and has searched directly and indirectly - for example through their effect on B physics -   for new particles that  are predicted in extensions of the standard model with a DM candidate. Matching the properties of the  Higgs particle imposes strong constraints on the parameter space of DM models. Furthermore the LHC is  searching for DM signals in events where a pair of DM particles are produced in association with a jet or a photon~\cite{Aad:2012fw,ATLAS:2012ky,Chatrchyan:2012me}, these searches provide the best limit on effective DM interactions with nucleons for light DM. 
Confronting the predictions from  a variety of particle physics models which possess
a stable weakly interacting massive particle  (WIMP) with all these observations will help to corner the properties  of a new cold dark matter candidate or show that one has to turn to other candidates that act rather as warm dark matter. 

The interpretation of the abundant recent and upcoming data from these different types of experiments requires  tools to compute
accurately the  various signals for  DM interaction and this in the context of different particle physics models. 
\micro~ is a code designed to predict dark matter signatures in various extensions of the SM which possess a discrete symmetry that guarantees the stability of the DM particle.\footnote{ DarkSUSY~\cite{Gondolo:2004sc}, Isared~\cite{Baer:2004qq} and SuperIsorelic~\cite{Arbey:2009gu} are other public codes that compute the signatures of dark matter.
These codes are confined to supersymmetric models.} 
 This code includes some of the common extensions of the SM : the minimal supersymmetric standard model (MSSM) and 
its extensions (NMSSM,CPVMSSM), the little Higgs model, the inert doublet model  and incorporates various 
constraints from collider physics and precision measurements for some of these models. 
Furthermore the code can accommodate user defined models. In this new version we extend the functionalities 
of \micro~ in three 
main directions: 
 
 \begin{itemize}
 \item{}  generalizing the Boltzmann equations to include asymmetric dark matter and  semi-annihilations. The DM asymmetry is taken into account when computing direct/indirect detection rates. 
 \item{}  incorporating loop-induced decays of Higgs particles to two-photons and two-gluons, and computing the signal strength for Higgs production in various channels that can be compared to results from LHC searches. Limits on the Higgs sector can also be obtained through an interface to HiggsBounds
  \item{}  adding a module for neutrino signature from DM capture in the Sun and the Earth 
 \end{itemize}

In addition the following improvements to the previous version are incorporated
\begin{itemize}
\item
Annihilation cross sections for  some selected 3-body processes in addition to 
the  2-body tree-level processes. The 3-body option can be included in the computation of the relic density and/or for annihilation of dark matter in the galaxy.
\item{}
Possibility of using different tables for the effective degrees of freedom ($g_{eff}(T), h_{eff}(T)$) in the early Universe
\item
Annihilation cross sections for  the loop induced  processes  
$\gamma\gamma$ and $\gamma Z^0$ in the NMSSM  and the CPVMSSM
\item{}
New function for incorporating DM clumps
\item{}
New function to define the strange quark content of the nucleon
\item{}
The LanHEP~\cite{Semenov:2008jy} source code for new models is included
\item{}
New models with scalar DM are included (Inert doublet model~\cite{Barbieri:2006dq} and model with $Z_3$ symmetry~\cite{Belanger:2012vp}) 
\item{}
New implementation of the NMSSM which uses the Higgs self-couplings and the particle spectrum calculated in
NMSSMTools\_4.0~\cite{Ellwanger:2005dv,Ellwanger:2006rn}
\item{}
New versions of spectrum generators used in the MSSM (Suspect$\_2.4.1$)~\cite{Djouadi:2002ze} and in the CPVMSSM (CPsuperH$2.3$)~\cite{Lee:2012wa}
\item{}
Extended routines for flavor physics in the MSSM 
\item{}
New facilities to compute DM observables independently of the model
\item{}
Update in interface tools to read files produced by other codes, this allows easy interface to other codes
\end{itemize}

\micro3 contains all the routines 
available in previous versions although the  format used to call some routines has
been changed. In particular global variables are used to specify the input parameters of various routines. 
All functions of \micro~ are described in the manual, manual3.pdf, 
to be found in the {\it man} directory, see also the lecture notes~\cite{TASI}.

In this paper we first describe in section 2 the generalization of the relic density routine to include asymmetric dark matter and semi-annihilation processes as well as more general thermodynamics of the Universe. Section 3 describes the new module to compute the neutrino-induced muon flux from DM captured in the Sun or the Earth. 
In section 4, we  give a brief description of improvements on the computation of the direct and indirect DM signals, in particular to take into account possible DM clumps. 
In section 5 we describe new facilities to compute more precisely the partial widths of the Higgs and    incorporate LHC results on the Higgs.  A brief description of SUSY models is provided in section 6. Section 7 contains a description of the functions that allow to work with external codes. 
Finally  sample results  are presented in Section 8.

\section{Generalization of the relic density computation}

The Boltzmann equation that describes the number density of DM  that is used to solve for the abundance of DM today and obtain its relic density is based on several assumptions. First, it is assumed that the DM candidate is either a self-conjugate particle or that there is no DM anti-DM asymmetry, and that there exists a discrete symmetry such as R-parity in supersymmetry that guarantees the stability of the lightest particle. Furthermore this symmetry permits only annihilation of pairs of DM particles into SM particles. Second, the annihilation cross section of DM particles is taken to be typically of electroweak strength such that DM particles are in thermal equilibrium in the early Universe and eventually freeze-out as the Universe cools down. Third, the contribution of the relativistic degrees of freedom to the entropy density and energy density of the Universe are taken to go as $T^4$ and $T^3$ respectively. 

In the following we will modify some of these assumptions by introducing the possibility of asymmetric DM and of new annihilation processes called semi-annihilation where two  DM particles annihilate into another DM particle and a SM particle. We will furthermore include the possibility of DM annihilation into 3-body processes. We will continue to assume that DM annihilation is typically of electroweak strength and keep the case of very weak interactions for further development.  We will also postpone the full implementation of a modification of the entropy and energy contributions to the Universe for a further version but will discuss the possibility to modify the temperature dependence of the effective number of degrees of freedom.

\subsection{Asymmetric dark matter}

We consider a dark matter particle that is not self-conjugate, with the particle being $\chi=\chi^+$ and the antiparticle $\bar\chi=\chi^-$. This occur for example when the DM particle is a Dirac fermion~\cite{Belanger:2007dx} or a complex scalar ~\cite{Barger:2008jx,Belanger:2012zr}. The Boltzmann equations for the abundance  of a DM particle --defined as the number density divided by the entropy density - have to be generalized~\cite{Graesser:2011wi,Iminniyaz:2011yp,Ellwanger:2012yg}. Recall that for a self-conjugate DM particle,$\chi$, 
\begin{equation}
\frac{dY_\chi}{ds} = \frac{<\sigma v> }{3H} \left( Y_\chi^2-Y_{eq}^2\right)
\label{eq:Y}
\end{equation}
where $<\sigma v>$ is the relativistic thermally averaged
annihilation cross-section, $Y_{eq}$ is the equilibrium abundance, $s$ the entropy and $H$  the Hubble constant.

We define $Y^+$ and $Y^-$ as the abundances of the DM particle/anti-particle ($\chi/\bar\chi$). 
At very high temperature, we assume that an asymmetry has already taken place such that $Y^+ \neq Y^-$ but that at later times the system is in equilibrium in a universe governed by a standard cosmology. 
We assume that annihilation to SM particles occurs only  between $\chi^+$ and $\chi^-$ and not among same charge DM. This could be the result of some conserved global charge carried by the DM particles. One should therefore expect that the difference between  the abundances $Y^+$ and $Y^-$  does not change during the thermal evolution. 
The evolution equation is then

\begin{equation}
  \frac{dY^\pm}{ds} = \frac{1}{3H} \left( <\sigma v^{+-}> (Y^+Y^--Y^+_{eq}Y^-_{eq})   \right)
  \label{eq:Ypm}
\end{equation}
with $Y^+_{eq}(T)=Y^-_{eq}(T)=1/2Y_{eq}(T)$ and 
 $\frac{1}{2}<\sigma v^{+-}>=\frac{1}{2}<\sigma v^{-+}>=<\sigma v>$.

 
 Subtracting the two equations in   Eq.~\ref{eq:Ypm} we obtain
\begin{equation}
\frac{dY^+}{ds} -\frac{dY^-}{ds}=0
\end{equation}
implying that the difference of abundances 
\begin{equation}
\label{deltaY}
    \Delta Y = Y^+ - Y^-
\end{equation}
 is constant as expected.

To proceed, we define the quantity  $Y\equiv 2\sqrt{Y^+ Y^-}$ which is equivalent to the abundance $Y=Y_\chi$ for a self-conjugate DM.  Using the abundance for each species, 
\begin{equation}
Y^\pm=\frac{\pm\Delta Y +\sqrt{Y^2+\Delta Y^2}}{2}
\end{equation}
the evolution equation for $Y$ is derived
\begin{eqnarray}
\label{eq:omega:asymm}
\frac{dY}{ds}&=&\frac{<\sigma v>}{3H} \frac{Y^++Y^-}{Y} \left(Y^2-Y_{eq}^2\right)\nonumber\\
&&=\frac{<\sigma v> }{3H}  \left(Y^2-Y_{eq}^2\right)\sqrt{1+\left(\frac{\Delta Y}{Y}\right)^2}
\end{eqnarray}
This equation is similar to the equation for a self-conjugate DM candidate save for an additional term. It can be solved numerically using the usual procedure~\cite{Belanger:2004yn}, to obtain the abundances today, $Y_0=Y(T=T_0)$.
The total relic density of the DM particle/antiparticle is then given by
\begin{equation}
\label{eq:Omega:asym}
\Omega h^2 =\frac{8\pi }{3 H^2_{100}}\frac{m_\chi}{M_{\rm Planck}}\frac{\sqrt{Y_0^2+\Delta Y^2}}{s_0} 
\end{equation}
where $s_0$ is the entropy today, $H_{100}=100 {\rm km/s/Mpc}$ (the Hubble constant today is $H_0=H_{100} h$), and $M_{\rm Planck}$ is the Planck mass.
The relic density of each specie can be expressed in terms of the asymmetry parameter $\delta_{DM}$,
\begin{equation}
\label{eq:omegaPM}
\Omega_{\pm} h^2 =\frac{\Omega h^2}{1+e^{\mp \delta_{DM}}}
\end{equation}
with
\begin{equation}
\label{eq:dDM} 
\delta_{DM} ={\rm log} \left( \frac{\sqrt{Y_0^2+\Delta Y^2}+\Delta Y}{\sqrt{Y_0^2+\Delta Y^2}-\Delta Y}\right).\end{equation}
Note that the asymmetry, $\Delta Y$,  will always increase the relic abundance as compared to the one obtained in the  same model but without an asymmetry.

\begin{figure}[t] 
   \centering
   \includegraphics[width=8cm]{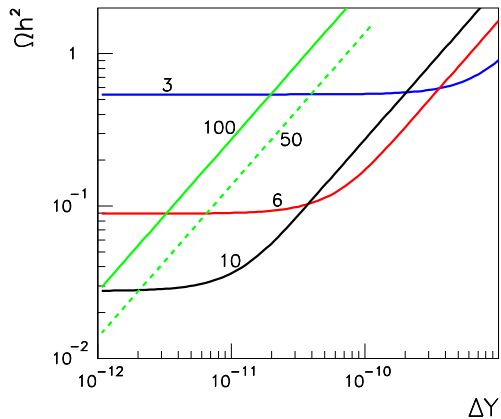}\quad
\vspace{-1cm}
   \caption{$\Omega h^2$ as a function of $\Delta Y$ for different DM masses in the 4th generation Dirac neutrino dark matter model for DM masses of $3,6,10,50,100$~{GeV} as indicated.}
   \label{fig:asym}
\end{figure}

To illustrate the impact of the asymmetry on the relic density, we use a simple extension of the SM with one extra lepton generation, $e_4,\nu_4$ where $\nu_4$ is a massive Dirac neutrino. These new leptons interact with SM gauge bosons but do not mix with the first  three generations. We assume an ad-hoc discrete symmetry where only $e_4,\nu_4$ have $R=-1$ while SM particles have $R=1$. Thus when $\nu_4$ is lighter than $e_4$ it is  stable and a potential DM candidate.
The dependence of $\Omega h^2$ on $\Delta Y$ for several DM masses is displayed in Fig.~\ref{fig:asym}.
\footnote{This model is only illustrative and would need to include new quarks to be anomaly free,  furthermore  the light DM masses are incompatible with the bounds on the Higgs invisible width that can be obtained from LHC results~\cite{Belanger:2013kya,Giardino:2013bma}.}
When $\Delta Y$ is small there is no changes in the symmetric component and the asymmetry will lead to a small increase in the relic abundance. When the asymmetry times $\sigma v$ is large, $Y_-$ is very small and the relic density is determined by $\Delta Y$, this shows up as a linear dependence of the relic density with $\Delta Y$ in Fig.~\ref{fig:asym}.  Note that since in this case one DM component has disappeared completely, a signal in indirect detection is not expected.
The linear dependence  starts at lower values of $\Delta Y$ for larger DM masses, since in this model $\sigma v$ also increases with the DM mass up to $m_{DM}\approx 50~{\rm GeV}$.
Note that when finding the starting point for the solution of  the evolution equation we assume that $Y-Y_{eq} \gg Y_{eq}$,  when $\Delta Y$ times $\sigma v$ is very large this assumption will fail, this means that  the freeze-out picture is inadequate.    
In this example $\sigma v$ is strongly enhanced when $m_{DM}\approx m_Z/2$ and our solution method breaks when $\Delta Y\approx 10^{-10}$ for this mass.

In ~\micro~the asymmetry  $\Delta Y$  is  a global parameter,   \verb|deltaY|,  that can be set 
anywhere in the code.  The relic density computation takes it into account automatically
to obtain both $\Omega h^2$ and  $\delta_{DM}$, Eq.~\ref{eq:dDM}. The latter  is also a global parameter {\tt DMasymm}, see Table (\ref{paramTab}).    Furthermore the  parameter {\tt DMasymm}  will have a
direct  impact on other observables computed in \micro.

The computation of the rate for dark matter scattering on nuclei is normally done averaging the cross section for
DM and anti-DM scattering on nucleons and depends on the local DM density which is the sum of DM ($\rho_{\chi}$) 
and anti-DM ($\rho_{\bar\chi}$)density. When the asymmetry parameter
$\delta_{DM}$ is non-zero,  the rate will  take into account the relative abundance of matter and anti-matter given 
the value of  the local DM density specified as a global parameter.
For indirect detection, the production rate of particles from DM annihilation is also modified to take into 
account the density of DM and anti-DM,  
\begin{equation}
Q= \frac{1}{2} \langle \sigma v\rangle\frac{\rho_\chi \rho_{\bar\chi}}{m_\chi^2 } \frac{dN_a}{dE},
\end{equation}
and the local density is $\rho_\chi+\rho_{\bar\chi}$.
Note that the parameter  \verb|DMasymm=|$\delta_{DM}$, which characterizes the asymmetry between the relic density of the two components, is also a global parameter. If reset after the relic density computation, it will be taken into account for computing direct and indirect detection rates and thus will allow the user to examine the impact of an asymmetry on these observables.




\subsection{Semi-annihilation}

With discrete symmetries larger than $Z_2$, one introduces new DM annihilation processes as well as the possibility of having more than one DM candidate corresponding to the lightest particle of two distinct dark sectors. Although \micro~ has been modified to include more than one DM candidate~\cite{Belanger:2012vp}, this possibility has been implemented only within the context of some specific DM models and is not included in this public distribution. In this section we therefore discuss only the impact of new processes of the type
 $\chi\chi\rightarrow \bar\chi X$ where $\chi$ is any particle from the dark sector ($\bar\chi$ correspond to anti-particles) and $X$ is any SM particle.
 These processes are called {\it semi-annihilation} ~\cite{Hambye:2008bq,D'Eramo:2010ep,Belanger:2012vp}
  
  The simplest example of a discrete symmetry which leads to semi-annihilation processes is  $Z_3$. The SM particles have $Z_3$ charge 0, while particles (anti-particles) in the dark sector have $Z_3$ charges $1$ and $-1$ respectively. Hence such models feature as usual only one DM candidate. 
With semi-annihilations, the equation for the number density reads
\begin{equation}
\frac{dn}{dt}=-\langle v\sigma^{\chi \bar\chi \rightarrow  XX} \rangle \left(n^2-{n_{\rm eq}}^2 \right) -\frac{1}{2} \langle v\sigma^{\chi\chi\rightarrow
\bar\chi X}  \rangle
\left(n^2-n \, {n_{\rm eq}} \right) -3H n.
\end{equation}
 We define 
\begin{equation}
\sigma_v\equiv  \langle v\sigma^{\chi \bar\chi \rightarrow XX}  \rangle+\frac{1}{2}\langle v\sigma^{\chi\chi\rightarrow
\bar\chi X}\rangle \quad {\rm and} \quad
\alpha=\frac{1}{2} \frac{\langle v\sigma^{\chi \chi\rightarrow \bar\chi X}\rangle}{\sigma_v},
\end{equation}
which means that $0\leq \alpha\leq 1$. In terms of the abundance, $Y=n/s,$ where $s$ is the entropy density, we obtain
\begin{equation}
\frac{dY}{dt}=-s \sigma_v \left(Y^2-\alpha Y {Y_{\rm eq}} -(1-\alpha) {Y_{\rm eq}}^2 \right) 
\end{equation}
or  \begin{equation}
3H\frac{dY}{ds}=\sigma_v \left(Y^2-\alpha Y {Y_{\rm eq}} -(1-\alpha) {Y_{\rm eq}}^2 \right).
\end{equation}
 To solve this equation we follow the usual procedure~\cite{Belanger:2004yn}. Writing $Y=Y_{\rm eq}+\delta Y$ we find the starting point for the numerical solution of this equation with the Runge-Kutta method using
\begin{equation}
3H\frac{d{Y_{\rm eq}}}{ds}=\sigma_v {Y_{\rm eq}}\delta Y \left( 2-\alpha \right),
\end{equation} 
where $\delta Y\ll Y$. This
is similar to the standard case except that  $\delta Y\rightarrow  \delta Y (1-\alpha/2)$. 
Furthermore, when solving numerically the evolution equation, the decoupling condition
$Y^2\gg {Y_{\rm eq}}^2$ is modified to 
\begin{equation}
Y^2\gg \alpha Y {Y_{\rm eq}} + (1-\alpha) {Y_{\rm eq}}^2.
\end{equation}
This implies that the freeze-out starts at an earlier time and lasts until a later time as compared with the standard case.
Although semi-annihilation processes can play a significant role in the computation 
of the relic density, the solution for the abundance depends only weakly on the parameter $\alpha$, typically   only by a few percent.
This means in particular that  the standard  freeze-out approximation works with a good precision. Explicit examples of the impact of semi-annihilation can be found in ~\cite{Belanger:2012vp,Belanger:2012zr}. The function {\tt onechannel} (see section~\ref{sec:relic:function}) can be used to obtain the contribution of each semi-annihilation channel and from there extract $\alpha$. 

There is no special need to indicate to \micro~ that a model contains  semi-annihilation processes, as long as all particles in the dark sector are properly identified, \micro~ will find the relevant processes. 
Note that the modified Boltzmann equation including both asymmetric DM and semi-annihilation is implemented in the code, however both terms cannot be present simultaneously in a specific model.

\subsection{Thermodynamics of the Universe}

The formulas that give the contribution  of one relativistic 
degree of freedom to the  entropy density and  energy  density of the Universe are
rather simple. For bosons,  
\begin{equation}
\rho_b=\frac{\pi^2}{30}T^4\;,\;\;\;s_b=\frac{2\pi^2}{45}T^3
\end{equation}
while he corresponding  functions for fermions have an additional factor $\frac{7}{8}$.
In the generic case one can write
\begin{equation}
\rho(T)=\frac{\pi^2}{30}T^4g_{eff}(T)\;,\;\;\;s(T)=\frac{2\pi^2}{45}T^3
h_{eff}(T)
\end{equation}
where $g_{eff}(T)$/$h_{eff}(T)$  count the effective numbers of degrees of
freedom.\footnote{$g_{eff}(T)$ and $h_{eff}(T)$ are related via the condition 
$\frac{d\rho(T)}{dT}=T\frac{ds(T)}{dT}$}   At freeze-out temperatures, the  contribution of DM  should be strongly suppressed, only
SM particles contribute to $g_{eff}$ and $h_{eff}$.\footnote{We assume that at low temperatures when DM 
contributes  significantly  to the matter density,  the formation of DM has been completed and $Y(T)={\rm Const}$.}
At low temperature $(T\ll 100 MeV)$, the  effective number of degree of freedom is 
defined by photons, neutrinos, light leptons.  As the temperature increases,  one 
has to add the quarks degrees of freedom. There is a problem of correct matching 
between the low and high temperature region. 

\begin{figure}
\begin{center}
\includegraphics[height=6cm, width=7.2cm]{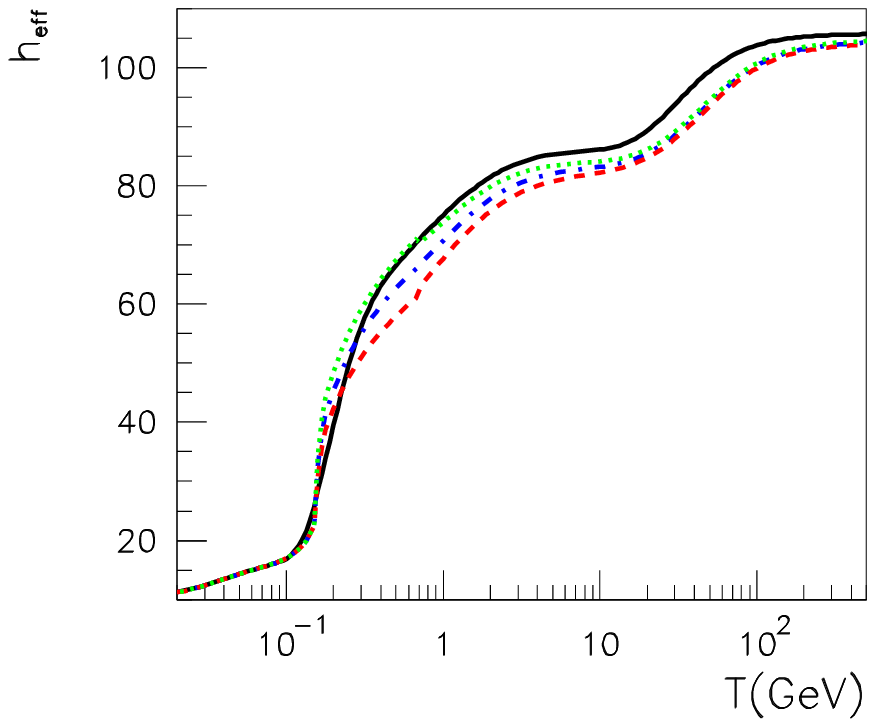}
\includegraphics[height=6.1cm, width=7.2cm]{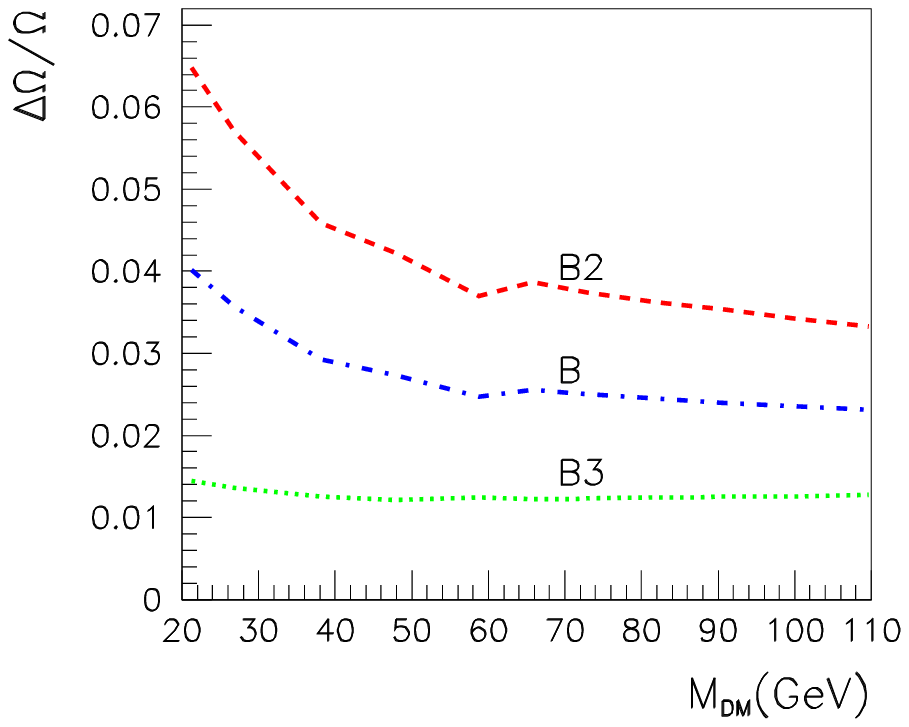}
 \vspace{-.8cm}
 \caption{ Left panel : Values of $h_{eff}$ as a function of T for different QCD equation of states, corresponding to the default \micro~ table with $T_{QCD}=150{\rm MeV}$ (black/full), and to case B (blue/dot-dash), B2(red/dash), B3(dot/green) of Ref.~\cite{Hindmarsh:2005ix}. Right panel : Relative difference between $\Omega h^2$  computed with the default table for $h_{eff}$ and cases B,B2,B3, same color code as left panel.  }
\label{fig:heff}
\end{center}
\end{figure}

 This issue was studied in~\cite{Hindmarsh:2005ix} where
other functions which feature different matching between the two regions  have been derived.
By default \micro~ uses the tabulated functions for $g_{eff}(T)/h_{eff}(T)$ 
obtained in \cite{Olive:1980wz,Srednicki:1988ce}, these tables assume that $T_{QCD}=150{\rm MeV}$. The temperature dependence of $h_{eff}$ is displayed in Fig.~\ref{fig:heff}.

The tabulated functions of ~\cite{Hindmarsh:2005ix} are also available. $h_{eff}(T)$ for three of those functions, B, B2 and B3, are compared to the default function in Fig.~\ref{fig:heff}, numerically the difference is small. A function ({\tt loadHeffGeff}) allows to substitute any table, see section~\ref{sec:relic:function}. 
The different functions for $h_{eff}$ can induce a shift in the value of the relic density, typically of a few percent. The larger shifts  are expected for light DM masses, especially for those where the freeze-out temperature is near the QCD transition region (that is $T_{fo}\approx m_{DM}/25 \approx 150{\rm MeV}$). As an example, we consider  the MSSM and display in Fig.~\ref{fig:heff} the resulting shift in $\Omega h^2$ for different choices of $h_{eff}(T)$  as a function of the neutralino mass. The neutralino mass is varied by changing  $M_1$ while keeping $M_{r3}=M_1+51{\rm GeV}$,
for each point the value of $\mu$ is adjusted such that $\Omega h^2\approx 0.1147$~\cite{Hinshaw:2012fq}. Other MSSM soft parameters are fixed to $M_2=400 GeV$, $M_3=1.8{\rm TeV}$, 
$M_{Li}=M_{Ri}=500{\rm GeV}$,$M_{Qi}=2{\rm TeV}$,$M_{ui}=M_{di}=1.5{\rm TeV}$, $A_t=A_b=-2.8{\rm TeV}$,
$M_A=700{\rm GeV}$, $\tan\beta=35$.

\subsection{Three-body processes}
\label{sec:three-body}

The pair annihilation of DM particles into a pair of SM particles may not be allowed for some kinematic and/or symmetry reasons. A known example is the suppression of the s-wave production of massless fermions in the case of a Majorana DM. This suppression is lifted by the emission of photons in association with the fermion pair. Another example to which we now turn is the case where annihilation into gauge bosons is kinematically  suppressed. The cross section for annihilation into off-shell W's can then be as large as that of other 2 body annihilation channels. 
This effect was shown to be particularly important in models such as the inert doublet model where the DM candidate is a scalar and has a standard SU(2) coupling to W's~\cite{Yaguna:2010hn,Honorez:2010re,Goudelis:2013uca}.

\micro~ has  a switch that allows to include three and four-body final states involving one/two off-shell gauge boson (W/Z). A similar procedure is also used for the decay of a Higgs into 3/4-body final states.
For DM annihilation or co-annihilation, we consider processes such as
\begin{equation}
\chi\chi' \rightarrow X V
\label{eq:virtual}
\end{equation}
where $V=W,Z$.  To take into account the process with an off-shell  $V$, we use CalcHEP to compute the 3-body process
\begin{equation}
\chi\chi'\rightarrow X l l'
\end{equation}
where $ll'$ is one leptonic decay channel of $V$. 
To obtain the total annihilation cross section into a virtual $V$, we 
divide this result by the branching ratio, $Br(V\rightarrow l l')$. In general for $l,l'$ we use leptons of the  first generation except when $X$ is a first generation lepton (this can happen in some co-annihilation processes).  In this case  we use  second generation leptons to avoid additional poles in matrix elements from unwanted diagrams that do not contain a virtual $V$.  For processes involving a virtual $Z$ we use  $l l'=\nu\bar\nu$ to avoid extra diagrams with virtual photons, even though those are expected to be small due to the large virtuality of the photon.

Processes with three-body final states can contain diagrams that do not involve a virtual $V$, our procedure for reconstructing the total 3-body cross section from only one specific final state would in  this case be spoiled. Ideally 
we would need to select only the diagrams with a virtual $V$ corresponding to the process in Eq.~\ref{eq:virtual}; this however can induce breaking of $SU(2)$ gauge invariance.  To solve this problem we select  diagrams where the total number of virtual vector bosons (W and Z) is at least 1, thus  ensuring a gauge invariant set of diagrams.
 
If $X$ is also a SM  vector boson then the procedure we have just described does not work very well  since one or both of the $V$ can be off-shell. First, note that  far below the threshold for $VV$ production, the total cross section is twice the one for $Vll'$ since each vector boson can be off-shell, while very near threshold the 3-body result should be close to the 2-body result.
We therefore need to compute the cross section for four-body final states. This is however computer-time consuming and would significantly slow down our code to a point where large scans of parameter space cannot be realized. We therefore use a trick to evaluate the 4-body process from the 3-body result after applying a K-factor. 
To obtain the K factor we assume  that the 4-body matrix element is proportional to 
\begin{eqnarray}
       M_{4-{\rm body}} 
       &&\propto \left[ (m_1m_2)^{2} +\frac{1}{8}\left(s-m_2^2-m_1^2 \right)^2 \right]  \nonumber\\  
       &&\frac{\Gamma_1}{\left( (m_1^2 -{m'_1}^2)^2 - \Gamma_1^2 {m'_{1}}^2\right)}  
         \frac{\Gamma_2}{\left( (m_2^2 -{m'_2}^2)^2 - \Gamma_2^2 {m'_{2}}^2\right)}
         \end{eqnarray}
   where $m_1,m_2$ are virtual masses and  $m'_1,m'_2$ are pole masses, $\Gamma_1,\Gamma_2$ are particle widths and $s=4m_\chi^2$. Note that this is the form of the matrix element for Higgs decay into 4-body final state through  two virtual gauge bosons.  
We integrate this matrix element with  $\Gamma_1=0$ in order to  simulate the  3-body result, then we integrate the same matrix element with the correct value for  $\Gamma_1$. The ratio  of these integrals  gives the K factor.  We have compared the value of the annihilation cross section at freeze-out obtained using the K-factor with the one obtained with the complete 4-body kinematics, in the MSSM for the benchmark point described below the difference is  less than 5\%.

 We consider only the contribution from virtual W's and Z's. The contribution from a virtual Higgs is expected to be small because the width is very small.  Contributions from virtual top quarks could also be important for DM masses below the top threshold~\cite{Yaguna:2010hn}, such  contributions will be included in a future upgrade.

\begin{figure}[t] 
   \centering
   \includegraphics[width=7cm]{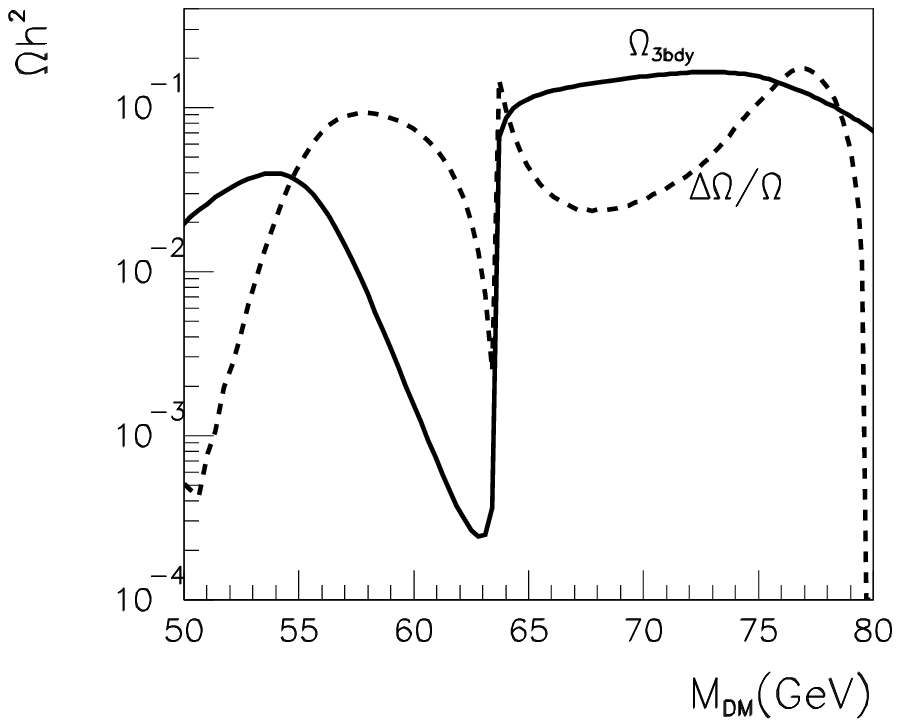}\quad
   \includegraphics[width=7cm]{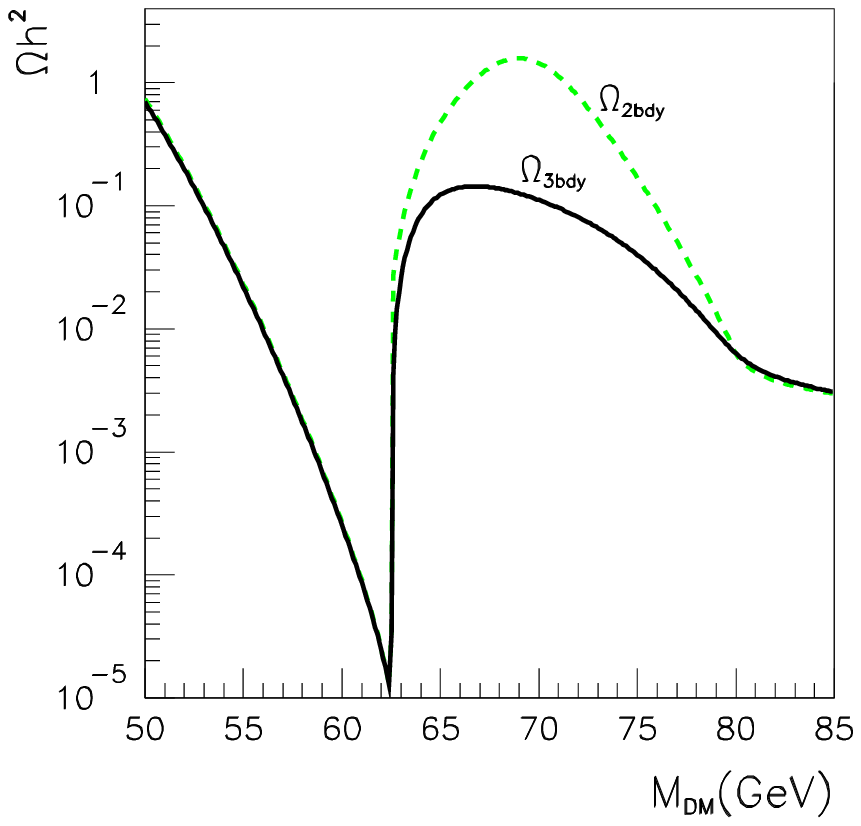}\quad
\vspace{-.5cm}
   \caption{Left panel : $\Omega h^2$ as a function of $M_{DM}$ in the MSSM  (full) and relative difference between the 3-body and 2-body value (dash). The soft parameters used are specified in the text.
   Right panel : $\Omega h^2$ as a function of $M_{DM}$ in the IDM  including  the 3-body final states (full)  and only 2-body final states (dash). }
   \label{fig:3bdy}
\end{figure}

In Fig.~\ref{fig:3bdy} we compare the DM relic density computed using 3-body final state with the one using only 2-body final  states. In the MSSM,  the LSP mass is changed by varying $M_1$, other soft parameters are fixed to 
$M_2=200 {\rm GeV}$, $M_3=1{\rm TeV}$, $\mu=126{\rm GeV}$, $M_A=1{\rm TeV}$, $\tan\beta=10$,
$M_{Li}=M_{Ri}=M_{Qi}=M_{ui}=M_{di}=1.5{\rm TeV}$,
 $A_t=2.8{\rm TeV}$.
Differences up to 17\% are found for masses just below the W-pair production. Furthermore differences up to 10\% are also found for LSP masses below $m_h/2$. Note however that in this example the relic density is very small for masses below $m_h/2$ since the Higgs exchange in s-channel gives an important contribution to DM annihilation. 
In the IDM, we take $\lambda_L=\lambda_2=0.01$, $m_h=125{\rm GeV}$, and vary the DM mass imposing 
$m_A-m_{DM}=m_{H^+}-m_{DM}=50{\rm GeV}$. For DM masses heavier than $m_h/2$ the difference between the 2-body approximation and the calculation that includes off-shell W/Z's can reach one order of magnitude, as was shown in Ref.~\cite{Honorez:2010re}.
For DM masses below $m_h/2$ where annihilation is completely dominated by Higgs exchange, we find basically the same result for the 2-body and 3-body final states, even though the Higgs width is quite different in the two cases. This is because the annihilation cross section does not depend on the value of the Higgs width as long as  the Higgs width and the DM pair annihilation are computed with the same approximation. Recall that in \micro~ the two-body option does not include the off-shell boson final states in the Higgs width, this explains the difference with the results in Ref.~\cite{Honorez:2010re}.  

\subsection{New functions and switches}
\label{sec:relic:function}

As in previous versions, the relic density computation is performed with the function \verb|darkOmega| as described in ~\cite{Belanger:2006is}, new features include the 3-body final state, which can be taken into account by specifying

\noindent
$\bullet$\verb|VWdecay,VZdecay|\\
Switches to turn on/off  processes with off-shell gauge bosons in final states for DM annihilation and particle decays.
If \verb|VW/VZdecay=1|, the  3-body final states will be computed for annihilation processes only while 
if \verb|VW/VZdecay=2| they will be included in coannihilation processes as well.
By  default  the switches are set to (\verb|VW/VZdecay=0|). \footnote{Including the 3-body final states can significantly increase the execution time for the relic density.}
Note that \micro~ calculates the width of each particle only once.  A second call to the function \verb|pWidth| (whether an explicit call or within the computation of a cross section)   will return the same result  even if the user has changed the {\tt VW/VZdecay} switch.  
To clean the decay table and re-activate the {\tt VW/VZdecay} switches,  one has to  call\\ 
$\bullet$\verb|clearDecayTable()| \\
to force \micro~ to recalculate particle widths according to the new value of {\tt VW/VZdecay}. 
The   {\tt sortOddParticles} command which must be used 
to recompute the particle spectrum after changing the model parameters also cleans  the decay table.

If  particle widths  were stored in the SLHA file downloaded by \micro, then the SLHA value will be used by \micro. To prevent downloading of decays one can download SLHA file using  {\tt slhaRead({\it fileName},mode=4)}, see description in Section \ref{ExtPackages}.

Two new  global parameters are available to set  the DM asymmetry:

\noindent
$\bullet$\verb|deltaY|\\
describes the difference between the DM and anti-DM abundances,  Eq. \ref{deltaY}.\\
\noindent
$\bullet$\verb|DMasymm|\\
is the relic density asymmetry parameter in Eq. (\ref{eq:dDM}) and is evaluated by \micro~ while calculating the relic density with an initial asymmetry  \verb|deltaY|. This parameter can also be reset  after the relic density computation and will be taken into account for direct and indirect detection rates.

The temperature dependence of the effective number of degrees of freedom can be set with

\noindent$\bullet$ \verb|loadHeffGeff(char*fname)|\\
loads the file \verb|fname| which contains a table of $h_{eff}(T), g_{eff}(T)$ . 
A positive  return value corresponds to the number of lines in the table. A negative return value indicates the line which creates a problem (e.g. wrong format), the routine returns zero when the file \verb|fname| cannot be opened. Several  files  containing tables of $h_{eff}(T), g_{eff}(T)$  are stored in the directory {\verb|sources/data|}.
 The default file is \verb|std_thg.tab|  ~\cite{Olive:1980wz,Srednicki:1988ce}   and is downloaded automatically if loadHeffGeff is not called by the user. Five other files \verb|HP_|{\it X}\verb|_thg.tab| where $X=\{A,B,B2,B3,C\} $ contain data sets from  ~\cite{Hindmarsh:2005ix}.
The user can substitute his/her own table as well, if so, the file must contain three columns containing the numerical values for $T$, $h_{eff}$, $g_{eff}$, the data file can also contain comments on lines starting with \verb|#|.

Several new functions that give information on the contribution of different channels  to DM annihilation have been added in this version

\noindent
$\bullet$ \verb|vSigma(T,Beps,fast)|\\
calculates the thermally averaged cross section for DM annihilation  times velocity  
at a  temperature T [GeV], see Eq.2.6 in ~\cite{Belanger:2001fz}. The value for $\sigma v$ 
is expressed in [pb].  The parameter \verb|Beps| defines the criteria for including coannihilation
channels. The \verb|fast=1/0| option switches between the {\it fast}/{\it accurate} calculation. 
The global array {\tt vSigmaTCh} contains the 
contribution of different channels to {\tt vSigma}. \verb|vSigmaTCh[i].weight| specifies the relative
weight of the $i^{th}$ channel \\
\verb|vSigmaTCh[i].prtcl[j]|  {\it (j=0, 4)}  define particles names for the $i^{th}$
channel.\\
The last record in \verb|vSigmaTCh| array has zero weight and 
NULL particle names.  In the Fortran version, the function 
 \verb|vSigmaTCh(i,weight,particles)| with $i\ge 1$, serves the same purpose.    This function returns 0
if $i$  exceeds the number of annihilation  channels and 1 otherwise.
 {\it real*8 weight} gives the relative contribution of each
annihilation channel. {\it character*10 particles(4)} contains the names of
particles in the  annihilation process.\\

\noindent
$\bullet$ \verb|oneChannel(Xf,Beps,p1,p2,p3,p4)|\\   
calculates the relative   contribution of the  channel \verb|p1,p2| $\to$ \verb|p3,p4|
to $(\Omega h^2)^{-1}$  where  \verb|p1,... p4| are particle names.  To make a
summation over several channels one can substitute  \verb|"*"| instead 
of the particle name.\\
\noindent
$\bullet$ \verb|omegaCh| \\
is an array that contains the relative contribution and particle names for each
annihilation channel. In the Fortran version one uses instead
the function\\
\noindent\verb|omegaCh(i,weight,particles)|. These arrays
are similar to {\tt vSigmaTCh} described above. The array {\tt omegaCh} is filled after calling either
{\tt darkOmegaFO} or {\tt printChannels}.


\section{DM in neutrino telescopes}
\label{neutrino}

After being captured, DM particles concentrate in the center of the Sun/Earth and 
then  annihilate into Standard Model particles. These SM particles further decay producing neutrinos that can be 
observed at the Earth.   Annihilation of DM particles in the galaxy is also a source of neutrinos. As for other indirect searches, the shape of the neutrino flux depends on the dominant DM annihilation channel into SM particles. Direct annihilation of DM into neutrinos gives  the hardest spectrum followed by muons and W's.  
As we will see below the signal for neutrinos coming from DM captured in a celestial body is essentially determined by the cross section for DM scattering on nuclei and is thus related to direct detection searches while galactic neutrinos probe the DM annihilation cross section. 

The neutrino  signal from DM captured in the Sun or Earth usually dominates over the one originating from DM annihilation in the galaxies. The best limit on galactic neutrinos have been achieved by IceCUBE~\cite{Abbasi:2012ws}, however the limit on the DM annihilation cross section are  still orders of magnitudes above the canonical cross section required to achieve the relic density assuming a standard cosmological scenario. 
Limits on the muon flux from neutrinos captured in the Sun were obtained by Super-Kamiokande, ~\cite{Tanaka:2011uf}, Baksan~\cite{Boliev:2013ai}, Amanda and ICECUBE~\cite{IceCube:2011aj},
and are currently more sensitive  than direct detection experiments in probing DM nuclei spin dependent cross sections on protons. 
The best limit on the SD cross section on protons is  currently obtained by ICECUBE, $\sigma^{SD}=1.2\times 10^{-40} {\rm cm}^{-2}$ for $m_{DM}=200{\rm GeV}$~\cite{IceCube:2011aj}.
 Note that ICECUBE with the DeepCore extension has a threshold energy of 10GeV~\cite{Ha:2012np}
 while Super-Kamiokande has a threshold of 1.6 GeV and is thus the best detector for light DM candidates. 
The neutrino flux and the induced muon flux from DM captured in the Sun and the Earth is a new module and will be described below. 

\subsection{Capture rate}\label{sec:neutrinos}

The capture rate for DM particles in the core of the Sun/Earth depends on the DM--nucleus scattering 
cross-section, as well as on the DM velocity distribution and local density. 
A DM particle  will get captured if after scattering with a nucleus at a distance  $r$ from the center, its velocity is lower than the escape velocity,
$v_{esc}(r)$, defined via gravitational potential
\begin{equation}
  \frac{v_{esc}(r)^2}{2}= \phi(r)
\end{equation}
 The capture rate was  derived in ~\cite{Gould:1987ir} 
 \begin{eqnarray}
{C_\chi}&=& \frac{\rho_\chi}{m_\chi}\int_0^\infty du f(u) \int 4\pi r^2 dr 
\sum_A \sigma_{\chi A} n_A(r) \frac{\beta_A}{\alpha_A} \theta( e^{-\alpha_A u^2}-e^{-\alpha_A( u^2+v^2_{esc}(r))/\beta_A})
\label{eq:cap}
 \end{eqnarray}
 where  $u$ is the WIMP velocity at infinity, $f(u)$ is the WIMP velocity distribution
normalized by the condition  
\begin{equation}
\label{f1_norm}
 \int_0^{\infty} u f(u) du = 1
\end{equation}
 $\sigma_{\chi A}$ is the scattering cross section for DM on nuclei, $n_A$ is the number density of nucleus $A$, and
  \begin{equation}
 \alpha_A=\frac{1}{3}  m_\chi m_A R_A^2 , \;\;\;    \beta_A=\frac{(m_\chi+m_A)^2}{4m_\chi m_A} \end{equation}
 where $m_A$ is the mass and $R_A$ the radius of nucleus A,
 The computation of $\sigma_{\chi A}$ takes into account the  fact that the WIMP scattering cross section is velocity dependent
which implies a form factor dependence  
 \begin{equation}
 |F(q^2)|^2=exp\left(\frac{-q^2 R_A^2}{3}\right)
 \end{equation}
where  $q$ is the momentum transfer.
For the Gaussian form factor\footnote{Note that in the direct detection module we use a Woods Saxon form factor rather than a Gaussian form factor as it is more accurate especially at large $q^2$,  however the Gaussian form factor can be integrated analytically.  }
 we assume that
\begin{equation}
R_A=\left(0.91 (m_{A}/{\rm GeV})^{1/3} +0.3\right) \times 10^{-13} {\rm cm} \end{equation}

For spin-independent interactions which add coherently, the sum is performed over all  nuclei up to $^{59}$Ni. For spin-dependent interaction, there is no coherence effect $\propto A^2$, therefore hydrogen gives the largest contribution, in this case we do not include the form factor. 
 The capture rate can be modified by finite temperature effects, however this effect is below 1\%~\cite{Gould:1987ir} so we neglect it. To calculate $n_A(r)$ we use the profiles 
given ~\cite{Asplund:2004eu} and \cite{Geochemistry}.

\subsection{Annihilation}
When the DM is  self-conjugated, the annihilation rate of captured particles is 
 $N_\chi^2 A_{\chi\chi}$ where $N_\chi$ is the number of DM particles and
 \begin{equation}
A_{\chi\chi}=  \frac{{\langle \sigma v \rangle}_{\chi\chi}}{V_{eff}} \;\;\;,
\label{eq:ann}
\end{equation}
$V_{eff}$ is the effective volume of DM in  the Sun/Earth. Assuming a Boltzmann thermal  distribution of   DM captured by the Sun/Earth one can get
\begin{equation}
V_{eff}= \left( 4\pi \int_0^{R} r^2 dr e^{-\frac{\phi(r)m_\chi}{T_\chi}}\right)^2/
   \left( 4\pi \int_0^{R} r^2 dr e^{-2\frac{\phi(r)m_\chi}{T_\chi}}\right)
\end{equation}
and $\phi(r)$ is the gravitational potential.
The WIMPs are trapped close to the center of the Sun, in this region we can consider that the temperature and the density are constant. We choose $T_\chi=T(\bar{r})$  where $\bar{r}$ is the mean radius of the WIMP distribution
\begin{equation}
\bar{r}=\left( \frac{6 T(\bar{r})}{\pi^2G\rho(\bar{r}) m_{\chi}}\right)^{1/2}
\label{eq:rbar}
\end{equation}
with $G$ the Newton's constant. The mean radius can be solved numerically from the temperature and density profile of the Sun/Earth. 

When  DM is not self-conjugated, an  additional term   $N_\chi N_{\bar{\chi}} A_{\chi\bar{\chi}}$  contributes to the DM annihilation rate where
 \begin{equation}
A_{\chi\bar{\chi}}=  \frac{{\langle \sigma v \rangle}_{\chi\bar{\chi}}}{V_{eff}}
\label{eq:ann2}
\end{equation}

\subsection{Evaporation}

DM captured in the Sun/Earth can scatter with nuclei inside the Sun/Earth and gain enough energy to escape from the Sun/Earth.
 Evaporation depends very sensitively on the DM mass, it is important for light DM.
  This effect was estimated in \cite{Gould:1987ju,Griest:1986yu}, approximate formulas for the evaporation rate were derived in ~\cite{Gould:1989tu}.
The evaporation rate ~\cite{Gould:1989tu}
\begin{equation}
E_\odot=\frac{8 \sigma_c}{\pi^3 \bar{r}^3} \bar{v} \frac{m_\chi \phi(\bar{r})}{T(\bar{r})} e^{-m_\chi \phi(\bar{r})/T(\bar{r})}
\label{eq:evap}
\end{equation}
  where 
$\bar{r}$ is  the mean radius (Eq.~ \ref{eq:rbar})
and $\bar{v}$ the mean speed of DM,
\begin{equation}
\bar{v}=\left( \frac{8 T(\bar{r})}{\pi m_{\chi}}\right)^{1/2}
\end{equation}
$T(r)$ is the temperature profile of the Sun/Earth.
The evaporation rate depends also on the integrated scattering cross section of DM on nuclei inside a radius $r_c$, $\sigma_c$,
\begin{equation}
\sigma_c=\int_0^{r_c} \sum_A  \sigma_{\chi A} n_A 4\pi r^2 dr
\end{equation}
where $r_c$ is taken to be the radius where the temperature drops to 95\% of its value, $T(r_c)=0.95 T(\bar{r})$.
Typically, evaporation in the Sun  affects significantly DM particles lighter than 3~GeV and is irrelevant 
for heavier DM particles.

\subsection{Neutrino flux}

When the DM is self-conjugate, the equation describing the evolution of the number of DM particles $N_\chi$ is simply
 \begin{eqnarray}
   \dot{N}_\chi& =& C_\chi -A_{\chi\chi} N_\chi^2-E_\chi N_\chi \,, 
\label{eq:ndot}
\end{eqnarray}
where the capture rate, the annihilation rate and the evaporation rate  were given above, Eqs.~\ref{eq:cap},\ref{eq:ann},\ref{eq:evap}.
When the evaporation is negligible and 
the capture and annihilation rates are sufficiently large, equilibrium is reached and the annihilation rate is only determined by the capture rate.
\begin{equation}
\label{eq:capture}
\Gamma_{\chi\chi}=\frac{1}{2} A_{\chi\chi} N_\chi^2= \frac{C_\chi}{2}.
\end{equation}
 The condition for equilibrium is $\sqrt{C_\chi A_{\chi\chi}} t \gg 1$. For the Sun, where $t_\odot=   4.57\times 10^9 {\rm yrs}$, equilibrium is easily reached, 
for the Earth, equilibrium might not be reached. 
To take into account the most general case, we do not make assumptions about equilibrium and solve Eq.~\ref{eq:ndot} numerically.

In models where the DM is not self-conjugate, one can get different capture rates  for particles and 
antiparticles. Furthermore both particle--particle (antiparticle--antiparticle) and particle--antiparticle 
annihilation channels exist. For example for a Dirac fermion DM one can get  $\lsp\lsp \rightarrow \nu\nu$ ($A_{\chi\chi}$) and 
$\lsp\bar\lsp \rightarrow X\bar{X}$ ($A_{\chi\bar\chi}$) where $X$ is any SM particle.
The equations describing the evolution of the number of DM (anti-)particles, $N_\chi(N_{\bar\chi})$, 
are then 
\begin{eqnarray}
   \dot{N}_\chi& =& C_\chi - A_{\chi\chi} N_\chi^2 -A_{\chi\bar\chi}N_\chi N_{\bar\chi} -E N_\chi\,, \nonumber\\
   \dot{N}_{\bar\chi}& =& C_{\bar\chi} - A_{\chi\bar\chi} N_\chi N_{\bar\chi} -A_{\bar\chi\bar\chi} N^2_{\bar\chi} -E N_{\bar\chi} \,,
\label{eq:ndot:asym}
\end{eqnarray}
The capture and annihilation rates can be evaluated as above, and the two coupled equations solved numerically to extract the number density of DM and anti-DM. In the case that evaporation is negligible and equilibrium is reached, these equations can be solved analytically, see  Appendix B.

After solving for the number density, one can compute the neutrino flux at the Earth.
First define,
\begin{eqnarray}
\label{eq:gammas}
\Gamma_{\chi\chi}&=&\frac{1}{2}A_{\chi\chi} N_{\chi}^2 \;;\;\; \Gamma_{\bar\chi\bar\chi}=\frac{1}{2}A_{\bar\chi\bar\chi} N_{\bar\chi}^2\;;\;\;
\Gamma_{\chi\bar\chi}=A_{\chi\bar\chi} N_{\chi} N_{\bar\chi}
\end{eqnarray}

The total neutrino spectrum at the Earth, assuming self-annihilation channels are solely into neutrino pairs,
is given by 
\begin{eqnarray}
  \label{eq:nu_spectra}
  \frac{d\phi_\nu}{dE_\nu} &=& \frac{1}{4\pi d^2} \left(
       \Gamma_{\chi\chi} Br_{\nu\nu}\, \frac{dN_{\nu\nu}}{dE} +
       \Gamma_{\chi\bar\chi} \sum_f Br_{f\bar{f}}\, \frac{dN_f}{dE} \right)\,, \nonumber\\
  \frac{d\phi_{\bar\nu}}{dE_{\bar\nu}} &=& \frac{1}{4\pi d^2} \left( 
       \Gamma_{\bar\chi\bar\chi} Br_{\bar\nu\bar\nu} \, \frac{dN_{\bar\nu\bar\nu}}{dE} + 
       \Gamma_{\chi\bar\chi} \sum_f Br_{f\bar{f}}\, \frac{dN_f}{dE} \right) \,,
\end{eqnarray}
where $d=d_{\odot}=1.5\times 10^8$~km for  the Sun  and
$d=d_{\oplus}=6.378$~km   for  the Earth,
$Br_{\nu\nu}$ is the branching fraction for annihilation into neutrino pairs, 
$Br_{f\bar{f}}$ the branching fraction into each particle/antiparticle final state $f\bar{f}$. 
$N_f$ and $N_{\nu\nu}(N_{\bar\nu\bar\nu})$ are the neutrino spectra resulting from 
those annihilations.
The neutrino spectra originating from different annihilation channels into SM particles and 
taking into account oscillations  and Sun/Earth medium effects were  computed in~\cite{Cirelli:2005gh}, we use the tables given 
there. Note that for the neutrino pair production an average over the three flavors in the annihilation process is assumed. 
In models where light DM annihilate directly into neutrinos of a given flavor (e.g. the light sneutrino scenario in ~\cite{Belanger:2010cd})
this is not the case, however this is still a good
approximation since almost perfect 3-generation mixing is expected for  neutrinos below 
10~GeV propagating in the Sun~\cite{Cirelli:2005gh}. 
 
\subsection{Muon fluxes}

Neutrinos that reach the Earth interact with rock below or water/ice within the detector and will give muons. 
To compare with the data, one must therefore compute the muon flux. First consider muons that are created inside the detector, leading to the contained muon flux, ~\cite{Erkoca:2009by}

\begin{equation}
\label{eq:contained}
   \frac{d\phi_\mu}{dE_\mu} =\rho_M N_A
\int\limits_{E_\mu}^{\mlsp} dE_{\nu}  \frac{d\phi_\nu}{dE_{\nu}}
\frac{d\sigma_{cc}(E_{\nu})}{dE_\mu}
\end{equation}
where $\rho_M$ is the water density, $N_A$ the Avogadro number, and $\sigma_{cc}$ the cross section for charged current interactions for neutrinos on nucleons, $\nu N\rightarrow \mu N'$. This cross section is averaged over protons and neutrons.

When neutrinos interact with matter outside the detector, one must also take into account the fact that muons lose energy on their way to the detector.
The energy loss for a muon is assumed to be
\begin{equation}
\frac{dE_\mu}{dz}= -\rho_M(\alpha +\beta E_\mu)
\end{equation}
where $\rho_M=2.6$~g/cm$^3$ is the rock density and  $\alpha=2\times 10^{-3}$~GeV\,cm$^2$/g,
$\beta=3\times 10^{-6}$~cm$^2$/g characterize the average energy loss of the muon 
traveling through rock or water. 
A muon of energy $E'_\mu$ will,  after propagating a distance $z$ have an energy $E_\mu$, where
\begin{equation}
    E'_{\mu}=  e^{\beta\rho z} \left(E_\mu +\frac{\alpha}{\beta}\right) -\frac{\alpha}{\beta}
\end{equation}
and its survival probability will be
\begin{equation}
  P_{surv}(E_\mu^{'},E_\mu) = \left(\frac{E_\mu}{E_\mu^{'}}\right)^y 
                                              \left(\frac{\alpha+\beta
E_\mu^{'}}{\alpha+\beta E_\mu}\right)^y
\end{equation}
where $y=m_\mu /(c\tau\alpha\rho_M)$ and $\tau$ is the muon lifetime.  


The upward muon flux is then~\cite{Erkoca:2009by}
\begin{equation}
\label{eq:upward}
   \frac{d\phi_\mu}{dE_\mu} = \rho_M N_A
\int\limits_{E_\mu}^{\mlsp} dE_{\nu}  \frac{d\phi_\nu}{dE_{\nu}}
  \int\limits_0^{z_{max}} dz 
 \frac{d\sigma_{cc}(E_{\nu})}{dE'_\mu} e^{z\beta\rho}
P_{surv}(E_\mu^{'},E_\mu) 
\end{equation}
where 
\begin{equation}
z_{max}= \frac{1}{\beta\rho_M} \log\left(\frac{\alpha+E_\nu\beta}{\alpha+E_\mu\beta}\right)
\end{equation}
Note that $E'_\mu(z=z_{max})=E_\nu$.

The event rate takes into account the effective area of the detector. For Super-K, 
the detector is cylindrical with a radius $R=18.9$\,m and a height $36.2$\,m. The muons that 
are stopped in the first 7\,m are not observed, this corresponds to the muon energy threshold 
of $E=1.6$ GeV. Furthermore, only muons of energy larger than $7.7$~GeV go through the 
detector. This means that for models with DM masses below 8 GeV, almost all 
the muons will be stopped within the detector. 


Because of oscillations,  the neutrino flux at the Earth will contain an almost equal mixture of  electron, muon  and tau neutrinos. Muon neutrinos will convert into muons detectable in a detector while electron neutrinos will produce electrons which will not propagate for very long. Finally tau neutrinos will produce taus which rapidly decay, one of the tau decay modes involves muons and will therefore induce a  correction to the $\nu_\mu$-induced muon flux. We do not take this effect into account.

\subsection{Functions for neutrino flux}
\label{neutrino_functions}
\noindent
$\bullet$ \verb| neutrinoFlux(f,forSun,nu, nu_bar)|\\
calculates  the muon neutrino/anti-neutrino  fluxes  near the surface of the Earth. 
This function  a)  calculates capture,   annihilation, and  evaporation rates,  Eqs.~\ref{eq:cap},\ref{eq:ann},\ref{eq:evap}; b)  solves numerically  Eq.~ \ref{eq:ndot} or
Eq.~\ref{eq:ndot:asym}; c) calculates all branchings of DM annihilation and substitute them in
Eq.~\ref{eq:nu_spectra} to get the fluxes.

Here  \verb|f(v)|  is  the DM velocity distribution   normalized  as in
Eq.~\ref{f1_norm}. 
The units  are $km/s$ for v and $s^2/km^2$ for  
f(v). At first approximation,   one can use the same  \verb|Maxwell| 
function introduced for direct detection.
   If {\tt forSun==0} then the flux of neutrinos from the Earth is calculated, otherwise this function computes the flux of neutrinos from the Sun as described in section~\ref{sec:neutrinos}.  The calculated fluxes are stored in {\tt nu} and {\tt nu\_bar}  arrays of dimension NZ=250.  
The neutrino fluxes are expressed in \verb|[1/Year/km|$^2$\verb|]|.
 The function
{\tt SpectdNdE(E,nu)} returns the differential flux of neutrinos in 
\verb|[1/Year/km|$^2$\verb|/GeV]|, and \\
\verb|     displaySpectrum(nu,"nu from Sun [1/Year/km^2/GeV]",Emin,Emax,1)|\\
allows to display the corresponding spectrum on the screen.

\noindent
$\bullet$ \verb|muonUpward(nu,Nu,rho, muon)|\\
calculates the muon flux which results from interactions of
neutrinos with rocks below the detector Eq.(\ref{eq:upward}). Here  {\tt nu} and {\tt Nu} are input arrays containing the
neutrino/anti-neutrino fluxes calculated by {\tt neutrinoFlux}. 
{\tt rho} is  the Earth density $\approx 2.6g/cm^3$. {\tt muon} is an
array which stores the resulting sum of $\mu^+$, $\mu^-$ fluxes. {\tt SpectdNdE(E,muon)}  gives the
differential muon flux  in \verb|[1/Year/km|$^2$\verb|/GeV]| units.

\noindent $\bullet$ \verb|muonContained(nu,Nu,rho, muon)|
calculates  the flux  of muons  produced in a given detector volume,
Eq.(\ref{eq:contained}).
This function  has the same parameters as \verb|muonUpward| 
except that the  outgoing  array gives the differential muon flux resulting from neutrinos converted to muons 
in a  $km^3$ volume given  in \verb|[1/Year/km|$^3$\verb|/GeV]| units.  \verb|rho| is the density of the detector in 
\verb|g/cm|$^3$.

Two functions allow to estimate the background from atmospheric neutrinos creating muons after interaction  with rocks below the detector  or with water inside the detector.
\noindent $\bullet$  \verb|ATMmuonUpward(cosFi,E)| calculates the sum of  muon
and anti-muon fluxes that  result from
interaction of  atmospheric  neutrinos with rocks. Units are  
\verb|[1/Year/km|$^2$\verb|/GeV/Sr]|. \verb|cosFi|  is the angle between the direction of
observation and  the center of the Earth. \verb|E| is the muon energy in
{\tt GeV}.\\
\noindent $\bullet$  \verb|ATMmuonContained(cosFi, E, rho)| calculates the flux
of muons caused by atmospheric  neutrinos  which are produced in a given  detector
volume in  \verb|[1/Year/km|$^3$\verb|/GeV]|. \verb|rho| is the
 density of the detector in \verb|[g/cm|$^3$\verb|]|. \verb|cosFi| and \verb|E| are the
same as above.

\section{Dark matter  detection}

The modules to compute the direct and indirect detection rates are described in Refs.~\cite{Belanger:2008sj,Belanger:2010gh}. Here we just mention the few upgrades  including improved values for the quark coefficient in nucleons relevant for direct detection, a new module to compute loop-induced annihilation of neutralinos into gauge bosons in the NMSSM and CPVMSSM and new functions for introducing DM clumps.

\subsection{Gamma-Ray line}

The annihilation of a pair of dark matter particles into photon pairs, or $\gamma Z$ gives a very distinctive DM signal, a monochromatic line at an energy $E=m_{DM}$ for $\gamma\gamma$ or $E=m_{DM}(1-m_Z^2/4m_{DM}^2)$  for $\gamma Z$.  
Possible evidence for such lines in FermiLAT data have been reported~\cite{Bringmann:2012vr,Weniger:2012tx} although the result is not yet confirmed by the FermiLAT collaboration and the DM origin of these lines remains to be established.
Such processes occur at the one-loop level and are therefore not computed in CalcHEP for generic models. However a module for
computing loop-induced neutralino pair annihilation processes in the MSSM ($\chi\chi\rightarrow \gamma\gamma,\gamma Z$ ) was provided in previous \micro~ versions~\cite{Boudjema:2005hb}.   
The code to compute the amplitudes was obtained with {\tt SloopS},  an automatic code for one-loop calculations in the context of the SM and the MSSM~\cite{Baro:2008bg,Baro:2009gn}.
In this version, a similar module is provided for the NMSSM~\cite{Chalons:2011ia} and the CPVMSSM using {\tt SloopS}.
The resulting routine,  {\tt lGamma.exe}, reads the [N/CPV]MSSM  parameters  from  an SLHA file, and computes the cross sections for
$\chi\chi\rightarrow \gamma\gamma,\gamma Z$.

The calculation of $\chi\chi\rightarrow \gamma\gamma,\gamma Z$ is based on the same functions for 
all MSSM-like models,   \\
\noindent
$\bullet$ \verb|int loopGamma(double *vcs_gz, double *vcs_gg)|\\
If the call was successful this function returns zero and assigns to  the variables \verb|vcs_gz|
and  \verb|vcs_gg| the cross sections  $\sigma v$ for $\gamma Z$ and $\gamma,\gamma$ in
$[cm^3/s]$.  The function \verb|gammaFlux| can then be used to calculate the photon fluxes, see example  in {\tt  main.c[F]}.
Note that to  activate the calculation of {\tt loopGamma}
one has to uncomment the statement\\
\verb|#define LoopGAMMA| \\
at the top of the file {\tt  main.c[F]}.

\subsection{Dark matter profile and clumps}

The indirect DM detection signals depend on the DM density in our Galaxy.
The DM density is given as the product of the local density at the Sun with the halo profile function 
\begin{equation}
\rho(r)=\rho_\odot F_{halo}(r)
\end{equation}
In \micro~ $\rho_\odot$ is a global parameter {\tt rhoDM} and  the Zhao profile~\cite{Zhao:1995cp}

\begin{equation}
\label{rho}
F_{halo}(r)=\left(\frac{ R_\odot}{r}\right)^{\gamma}
\left(\frac{r_c^{\alpha}+ R_\odot^{\alpha}}
{r_c^{\alpha}+r^{\alpha}}\right)^{\frac{\beta -\gamma}{\alpha}}
\end{equation}
with   $\alpha=1,\beta=3,\gamma=1, rc=20[kpc]$ is used by
default. $R_\odot$, the distance from the Sun to the galactic center,  is also a global parameter, {\tt Rsun}. The parameters of the Zhao profile  can be reset by\\ 
\noindent
$\bullet$ \verb|setProfileZhao(|$\alpha$,$\beta$,$\gamma$,{\it rc}\verb|)|\\
The functions to set another  density profile reads\\ 
\noindent
$\bullet$ \verb|setHaloProfile(|$F_{halo}(r)$\verb|)|\\
where $F_{halo}(r)$ is any function from distance from Galactic center,  $r$
has to    be   defined   in [kpc] units.
For instance, \verb|setHaloProfile(hProfileEinasto)|  sets Einasto profile\\
$$
F_{halo}(r)=exp\left[-\frac{2}{\alpha}\left(\left(\frac{r}{R_\odot}\right)^{\alpha}-1\right)\right]
$$
where by default $\alpha=0.17$,   but can be changed by \\ 
\noindent
$\bullet$ \verb|setProfileEinasto(|$\alpha$\verb|)| \\
The command \verb|setHaloProfile(hProfileZhao)| sets back Zhao profile. Note
that both {\tt setProfileZhao} and {\tt setProfileEinasto} call
{\tt setHaloProfile} to install the corresponding profile.

Dark matter annihilation in the Galaxy depends on the average of the square of the DM density, $<\rho^2>$. This quantity 
can be significantly larger than $<\rho>^2$ when clumps of DM are present~\cite{Lavalle:2006vb}.  
In \micro~,   we use  a simple model where $f_{cl}$ is a constant 
that characterizes the fraction of the total density due to clumps
 and  where all clumps occupy  the
same volume $V_{cl}$ and have a constant density $\rho_{cl}$. Then assuming  clumps do not  overlap, we get 
\begin{equation} 
    <\rho^2> = \rho^2 +  f_{cl}\rho_{cl}\rho .
\end{equation}
This simple description allows to  demonstrate  the main effect of clumps:  far from the Galactic center the rate of DM annihilation falls as $\rho(r)$ rather than as
$\rho(r)^2$. The parameters $\rho_{cl}$  and  $f_{cl}$ have zero default values.  
The routine to change these values is \\
 \noindent $\bullet$ \verb|setClumpConst(|$f_{cl}$,$\rho_{cl}$\verb|)| \\
To be more general, one could assume that $\rho_{cl}$  and  $f_{cl}$  depend on the distance from galactic center. The effect of clumping  is then described  by the equation 
\begin{equation}
<\rho^2>(r)=\rho(r)(\rho(r) +  \rho_{clump}^{eff}(r))
\end{equation}
and the function \\
\noindent $\bullet$ \verb|setRhoClumps(|$\rho_{clump}^{eff}$\verb|)|\\
allows to implement a more sophisticated clump  structure.

\subsection{Direct detection}
The computation of the direct detection rate was presented in ~\cite{Belanger:2008sj}.
One important theoretical uncertainty in  the direct detection rate comes from the quark coefficients in the nucleon, and in particular the strange quark content. For SI interactions the operator $\langle N|m_q \bar{\psi_q}\psi_q|N\rangle$ is interpreted as the contribution of the quark $q$ to the nucleon  mass, $M_N$ with 
\begin{equation}
\langle N|m_q \bar{\psi_q}\psi_q|N\rangle =f_q^N M_N
\end{equation}
\noindent
We define the operators
\begin{eqnarray}
\sigma_{\pi N}=m_l\langle p|\bar{u}u+\bar{d}d|p\rangle\\
\sigma_{s}=m_s\langle p|\bar{s}s |p\rangle
\end{eqnarray}
where $m_l=(m_u+m_d)/2$. The relative importance of strange quark in nucleon is described by the ratio 
\begin{equation}
y= \frac{2\langle p|\bar{s}s|p\rangle}{\langle p|\bar{u}u+\bar{d}d|p\rangle} =\frac{m_l}{m_s} \frac{2\sigma_s}{\sigma_{\pi N}}
\end{equation}

In earlier \micro~ versions, the strange quark content was extracted using  the operator
$\sigma_{0}=m_l\langle p|\bar{u}u+\bar{d}d -2\bar{s}s |p\rangle$, this  method  leads to large uncertainties in the strange quark content and in the direct detection rate~\cite{Bottino:1999ei,Ellis:2008hf}. 
Recent lattice QCD calculations have allowed a direct determination of $\sigma_s$ (see ~\cite{Thomas:2012tg}), thus reducing significantly the large uncertainty in the estimation of $y$, furthermore the central value for 
$\sigma_s$ is also rather small.
A compilation of recent lattice results is presented in  Table~\ref{lattice_DD}. Taking the weighted mean of these values
(${\sum x_i/\sigma_i^2}/{\sum 1/\sigma_i^2}$)  we get  
\begin{equation}
\label{eq:lattice}
\sigma_s=42\pm 5~{\rm MeV} \;\;\;   {\rm and } \;\;\; \sigma_{\pi N}=34\pm 2~{\rm MeV}. 
\end{equation}
Considering the better and more direct determination of $\sigma_s$ it is more convenient to set the quark from factor from $\sigma_s$ rather than from $\sigma_0$, other parameters are $m_u/m_d,m_s/m_d,\sigma_{\pi N}$. The quark coefficients are expressed as
\begin{equation}
 f_d^p = \frac{2\sigma_{\pi N}}{\left(1+\frac{m_u}{m_d}\right) m_p (1+\alpha)}  \;\;\;   f_u^p =\frac{m_u}{m_d} \alpha f_d^p\;\;\; f_s^p=\frac{\sigma_s}{m_p}
\end{equation}
where 
\begin{equation}
\alpha=\frac{\langle N|\bar{u}u|N\rangle}{\langle N|\bar{d}d|N\rangle} =\frac{2z-(z-1)y}{2+(z-1)y} \;\;\; {\rm and} \;\;\;
z=\frac{\langle N|\bar{u}u|N\rangle-\langle N|\bar{s}s|N\rangle}{\langle N|\bar{d}d|N\rangle-\langle N|\bar{s}s|N\rangle}\approx 1.49
\end{equation}
A new function to compute the quark coefficients is defined,\\ 
\noindent$\bullet$
{\tt calcScalarQuarkFF}($m_u/m_d$,$m_s/m_d$,$\sigma_{\pi N}$,$\sigma_s$)\\
computes the scalar coefficients for the quark content in the nucleon from the quark mass ratios
$m_u/m_d, m_s/m_d$ as well as from $\sigma_{\pi N}$ and $\sigma_s$.
The default values given in Table  \ref{FFTab} in the Appendix are obtained from 
the central values $\sigma_s=42{\rm MeV},\sigma_{\pi N}=34{\rm MeV}$ in Eq.~\ref{eq:lattice} and from $m_u/m_d=0.56, m_s/m_d=20.2$ ~\cite{Beringer:1900zz}.
Note that the  mass ratios  estimated from lattice results rather than from chiral perturbation theory are somewhat different $m_u/m_d=0.46\pm 0.05, m_s/m_d=27.5\pm 0.3$~\cite{Beringer:1900zz}.

\begin{table}[th]
 \caption{Nucleon coefficients}
 \label{lattice_DD}
\begin{center}
\begin{tabular}{|c|c|l|}
\hline
 $\sigma_s$  (MeV)  & $\sigma_{\pi N}$(MeV) & Reference \\   \hline
$8\pm 21$ &     & ~\cite{Oksuzian:2012rzb}\\\hline
$43\pm 10$ &  & ~\cite{Engelhardt:2012gd}\\\hline
$54\pm 8$ &  & ~\cite{Freeman:2012ry}\\\hline
$22\pm 37$ &  & ~\cite{Oksuzian:2012rzb}\\\hline
$31\pm 16$ & $50\pm 10$ & ~\cite{Young:2009zb}\\\hline
$34^{+31}_{-27}$ & $42^{+21}_{-6}$ & ~\cite{Durr:2011mp}\\\hline
$70\pm 68$ & $31\pm 5$ & ~\cite{Horsley:2011wr}\\\hline
$22\pm 20$ & $32\pm 2$ & ~\cite{Semke:2012gs}\\\hline
$21^{+44}_{-6}$ & $45\pm 6$ & ~\cite{Shanahan:2012wh}\\\hline
$125\pm 59$ & $43\pm 6$ & ~\cite{Ren:2012aj}\\\hline
$50\pm 18$ & & ~\cite{Junnarkar:2013ac}\\\hline
& $37\pm 10$ & ~\cite{Bali:2012qs}\\\hline
\hline
\end{tabular}
\end{center}
\end{table}

\section{The Higgs sector at  colliders}
\label{sec:higgs}

The LEP, Tevatron and LHC colliders have performed extensive searches for the Higgs boson(s).
Recently the LHC has announced the discovery of a Higgs-like particle with a mass $m_h=125.5 {\rm GeV}$.~\cite{Aad:2012tfa,Chatrchyan:2012ufa}  A signal was found notably in the two-photon channel and in the two gauge bosons decay modes. The observation of  this new particle  as well as the non-observation of a Higgs particle in many other channels at the LHC, LEP or Tevatron both put strong constraints on the parameter space of the extensions of the standard model.  
Two new features of \micro~ allow to compute more precisely the partial widths of the Higgs: 
 the partial width for the Higgs into off-shell  W/Z's are included following the approach  discussed in section~\ref{sec:three-body} and loop-induced couplings of the Higgs are included via effective vertices.

\subsection{Loop-induced Higgs production and decay}
\label{sec:loop}

The loop-induced couplings  of the Higgs to gluons and photons play and important role in  Higgs production and decay at colliders. Thus the \micro~ models which are based on tree-level vertices need to be extended to include these interactions.
For this we introduce   effective operators of the form $\lambda h F_{\mu\nu} F^{\mu\nu}$ for the CP even part  and
$\lambda' h F_{\mu\nu}$  $\tilde{F}^{\mu\nu}$ for the CP-odd part\footnote{These operators are also used in CalCHEP~\cite{Belyaev:2012qa}}. These operators receive
contributions from charged scalars, vector bosons or fermions. For a generic Lagrangian describing the Higgs interactions with fermions $\psi$, scalars $\phi$ and vector bosons $V^\mu$,
\begin{equation}
{\cal L}= g_{h\psi\psi} \bar\psi\psi h + i g'_{h\psi\psi} \bar\psi\gamma_5\psi h + g_{h\phi\phi} M_\phi  h\phi\phi +g_{hVV} M_V h V_\mu V^\mu
\end{equation}
the contribution to   the effective operators read
\begin{eqnarray}
\lambda &= & \frac{\alpha}{8\pi} \left[ g_{h\psi\psi} f_\psi^c q_\psi^2 \frac{1}{M_\psi} A_{1/2}(\frac{M_h^2}{4M_\psi^2})
- g_{hVV} f_V^c q_V^2 \frac{1}{2 M_V} A_{1}(\frac{M_h^2}{4M_V^2})\right.\nonumber\\
&+& \left. g_{h\phi\phi} f_\phi^c q_\phi^2 \frac{1}{2M_\phi} A_{0}(\frac{M_h^2}{4M_\phi^2})\right] 
\label{eq:lambda}
\end{eqnarray}
and
\begin{equation}
\lambda' =  \frac{\alpha}{16\pi} g'_{h\psi\psi} f_\psi^c q_\psi^2 \frac{1}{M_\psi} \tilde{A}_{1/2}(\frac{M_h^2}{4M_\psi^2})
\end{equation}
where q is the electric charge of the virtual particle for $\gamma\gamma$ operators and is 1 for the gluon operator, $f^c$ is the color factor associated with the virtual particle. For the fundamental SU(3) representation $f^c=3$ for $\gamma\gamma$ and $f^c=1/2$ for $gg$. For the adjoint representation $f^c=8$ for photons  and $f^c=-2$ for gluons. The functions $A_{1/2},A_1,A_0,\tilde{A}_{1/2}$ are defined in 
~\cite{Djouadi:2005gi,Djouadi:2005gj}. 

There are important QCD corrections to the effective vertices with coloured particles in the loop. For $h\gamma\gamma$ the QCD corrections are described as an overall factor
\begin{equation}
1+\frac{\alpha_s(M_h/2)}{\pi} C_l\left( \frac{M_h^2}{4M_l^2}\right)
\label{eq:hgam_qcd}
\end{equation}
where $M_l$ is the mass of the particle in the loop and the $C_l$ functions are known for fermions and scalars in the fundamental representation of SU(3). The code \verb|HDECAY|~\cite{Djouadi:1997yw}  is used to generate tables for these functions.

The  QCD corrections to $hgg$ to order $\alpha_s^3$   have been computed in the limit of a heavy top quark. These  include both vertex corrections  and corrections from gluon emission~\cite{Baikov:2006ch}.  The corrections are therefore split into a radiation factor and a vertex factor.
For lighter quarks or for new colored particles, only NLO vertex corrections are known.
To take into account QCD corrections for all colored fermions and scalars in the loop, we use the following prescription. 
We use a common radiation factor, $R$, for all colored particles and introduce additional  correction factors in the amplitude  for top/heavy new quarks ($C_t$), light quarks ($C_q$) or colored scalars ($C_s$). 
The $hgg$ vertex including QCD corrections  for SM fermions and colored scalars (such as squarks in the MSSM) thus reads
\begin{equation}
\lambda_{\rm QCD}=  - R 
\left[ A^{LO}_{htt} C_t + (A^{LO}_{hbb}+A^{LO}_{hcc}) C_q
+ \sum_{\tilde{q}} A^{LO}_{h\tilde{q}\tilde{q}}C_{\tilde q} \right],
\label{eq:hgg_qcd}
\end{equation}
where $A^{LO}_{hXX}$ denote the contribution of particle $X$ to Eq.~\ref{eq:lambda}.
The QCD factors  with $a= \alpha_s(Q)/\pi$, read ~\cite{Baikov:2006ch},
 \begin{equation}
R^2= 1+\frac{149}{12} a + 68.648 a^2 -212.447 a^3,
\end{equation}
and
 \begin{eqnarray}
 C_t&= &1+\frac{11}{4} a +(6.1537-2.8542 \lnTop)a^2\nonumber\\
 & +& (10.999-17.93\lnTop+5.47\lnTop^2)a^3 \\
 C_q&=& 1+\frac{11}{4} a \\
 C_{\tilde q}&=& 1+\frac{9}{2} a ,
 \end{eqnarray}
where   $Q=M_h/2$.
The top contribution  to the $hgg$ partial widths agrees with Ref.~\cite{Baikov:2006ch} to order $\alpha_s^3$. 
 Our approach which consists in a modification of the effective $h\gamma\gamma$ and $hgg$ vertices, necessarily incorporates only part of  the higher order corrections to the partial widths  (for example  for lighter quarks the contribution to order $\alpha_s$ are taken into account while  only some of the $\alpha^2_s$ contributions are included). 
Furthermore with the effective vertex approach we cannot introduce QCD corrections for the  interference term between two contributions.
It is therefore expected that we will have differences with other treatment of QCD corrections.  We indeed observe some differences with {\tt HDECAY}.
For instance for lighter quarks or for new particles, the QCD factor in the partial width is in agreement with {\tt HDECAY} only to order $\alpha_s$. Nevertheless 
for a SM  Higgs of 126 GeV, differences with {\tt HDECAY} are below the percent level.
For  the SM-like Higgs of the MSSM,  differences are also generally small (at the few percent level).
Larger discrepancies can be found for the heavier Higgs of the MSSM, in particular  when the parameters are chosen such that  there is a strong interference between the top and stop contribution. 
In supersymmetric models, higher-order SUSY corrections, the so-called $\Delta m_b$ corrections,  are taken into account by a redefinition of the $h\bar{b}b$ vertex. These are also included in ~\micro~ when defining the b-quark contribution to $h\gamma\gamma$ and $hgg$, such corrections are not included in the loop-induced processes in {\tt HDECAY} thus leading to discrepancies between the two codes when the b-quark contribution is important.

In most extensions of the standard model, the Higgs sector contains several doublets and/or singlets. 
For all Higgs scalars and pseudoscalars, we follow the same procedure for computing the $H\gamma\gamma$ and $Hgg$ vertices, including QCD corrections.
Note however that these corrections were obtained in the heavy top quark limit which would not be appropriate for the Higgses that are very heavy. Nevertheless we use these results for all masses, since a high precision is not essential for the loop-induced processes when the Higgs is very heavy, indeed  both the gluon fusion Higgs production and the decay in the  $\gamma\gamma$ mode are suppressed at large masses. 

The functions that can be used to compute the effective Higgs vertices are\\

\noindent 
$\bullet$\verb|HggF(Q),HggS(Q),HggV(Q)|\\
to compute the fermion/scalar/vector loop contribution to the $hgg,h\gamma\gamma$ vertices, $Q=(M_h/2M)^2$ where $M$ is the mass of the fermion/scalar/vector in the loop.

\noindent
$\bullet$
\verb|HggA(Q)|\\
to compute the fermion loop contribution to the $Agg,A\gamma\gamma$ vertices, $Q$ is defined as above.

\noindent
$\bullet$
\verb|Hgam1F(Q),Hgam1S(Q)|\\
QCD corrections to the fermion/scalar contribution to the $h\gamma\gamma$ effective vertex, see Eq.~\ref{eq:hgam_qcd}.

\noindent
$\bullet$
\verb|Hgam1A(Q)|\\
QCD corrections to the fermion contribution to the $A\gamma\gamma$ effective vertex, see Eq.~\ref{eq:hgam_qcd}.

Implementation of the loop-induced vertices can be realised using LanHEP, the necessary functions are described next.

\subsection{Implementation of effective vertices}
\label{sec:lanhep}

A new feature in LanHEP allows to extract the expression from any generated vertex. This can be used to implement the different contributions to the loop-induced $H\gamma\gamma$ and $Hgg$ vertices. 
The function which allows to extract a specific coupling is\\
 {\tt CoefVrt([ \sl particles \tt  ],[ \sl 
options \tt ])},\\
 where  {\sl particles} is the comma-separated list
of particles in the vertex, and {\sl options} specifies the tensor 
structure of the required coefficient as well as other options.  The only tensor structure that can be extracted are $1,\gamma^5,\gamma^\mu,\gamma^\mu\gamma^5, p^\mu $  and the function {\tt CoefVrt} must be called separately for each tensor structure. The second list can contain
the following symbols in any order:
\begin{itemize}
\item {\tt gamma} corresponds to only $\gamma^\mu$ matrices in the vertex, {\tt gamma5} to only  $\gamma^5$ matrices and 
{\tt gamma}, {\tt gamma5} to product of  $\gamma^\mu \gamma^5$ matrices. If none of these symbols are listed, then only 
terms which do not include $\gamma$-matrices will be extracted;

\item  {\tt moment( \sl P \tt )} where {\sl P} is a particle name selects 
the terms depending on this particle momentum, note that vertices corresponding to higher order operators and containing  the product of several momenta cannot be handled;
\item {\tt re, im} - take real or imaginary part of the expression. 
It can be combined with options above.
\item {\tt abbr} - use abbreviations in the returned expressions, this option is useful for long expressions.
\end{itemize}

The following example illustrates how LanHEP can be used to compute the sleptons contribution to the 
$h\rightarrow\gamma\gamma$ effective vertex in supersymmetry. First, we define terms depending on the 
loop functions :
\begin{quote} \tt \_p=[eL,eR,mL,mR,l1,l2] in
    parameter rhS\_\_p=creal(HggS((Q/2/MS\_p)**2)),
	          ihS\_\_p=cimag(HggS((Q/2/MS\_p)**2)).
\end{quote}
Here eL,...l2 are the names for the three generations of charged sleptons,
MSeL...MSl2 are their masses, and we declare the new parameters
rhS\_eL...ihS\_l2. The function {\tt HggS} describes the loop integrals in Eq.~\ref{eq:lambda}.
Then we obtain expressions from the vertices 
of the Higgs boson interaction with these sleptons (divided by the square of the slepton mass):
\begin{quote} \tt \_p=[eL,eR,mL,mR,l1,l2] in
    parameter AhS\_\_p=CoefVrt([anti(\_p),\_p,h]) /(mass \_p)**2/2.
\end{quote}
Note that in general it is preferable to specify whether one wants to extract the real or imaginary part of the vertex, this option was omitted here because we know that  this vertex is real. Similarly the option {\tt abbr} is useful to get a compact expression for the vertices involving staus.  
The charge of particles can, for example,  be defined  as
\begin{quote}
\tt \_p in [eL,eR,mL,mR,l1,l2] alias Q(\_p)=-1
\end{quote}
Finally, the contribution to the Higgs decay amplitude reads:
\begin{quote} \tt LAAhs=cabs((rhS\_\#sl+i*ihS\_\#sl)*AhS\_\#sl*Q(sl)*Q(sl) \\
\phantom{   xxxxxxxxx            }where    sl=[eL,eR,mL,mR,l1,l2]).
\end{quote}
In this statement the inner {\tt where } keyword is used to make the sum
of expressions where the list of symbols on  the right-hand side is substituted
into the template expression on the left-hand side. The  \# sign means concatenation
of the symbols to the left-hand and right-hand side. To this expression, one must in general add the relevant  color factor that appear in Eq.~\ref{eq:lambda} as well as the QCD factors in Eq.~\ref{eq:hgg_qcd}.

\subsection{Higgs  couplings and Higgs exclusion}

In addition to a measurement of the Higgs mass, the LHC collaborations provide  estimates of the signal strength ($\mu$) as compared to the SM expectation in various  channels. For example for the two-photon decay of a Higgs produced via gluon  fusion(ggF)
or vector boson fusion(VBF) 
\begin{equation}
\mu_{\gamma\gamma}^{ggF}=\frac{\sigma(gg\rightarrow h)_{NP} Br(h\rightarrow \gamma\gamma)_{NP}}{\sigma(gg\rightarrow h)_{SM} Br(h\rightarrow \gamma\gamma)_{SM}} \;\;\; \,u_{\gamma\gamma}^{VBF}=\frac{\sigma(WW\rightarrow h)_{NP} Br(h\rightarrow \gamma\gamma)_{NP}}{\sigma(WW\rightarrow h)_{SM} Br(h\rightarrow \gamma\gamma)_{SM}}
\end{equation}
where NP stands for the new physics model under consideration.  
The ratio of the Higgs partial widths can be obtained  easily from the coupling squared normalized to the SM coupling
\begin{equation}
c_{hXX}=g^2_{hXX}(NP)/g^2_{hXX}(SM)
\end{equation}
and similarly for loop-induced processes, 
\begin{equation}
c_{hGG}=\lambda^2_{hGG}(NP)/\lambda^2_{hGG}(SM)
\end{equation}
where $G=g,\gamma$ and  $\lambda_{hGG}$ are  given in Eqs.~\ref{eq:lambda},\ref{eq:hgam_qcd},\ref{eq:hgg_qcd}.
We make the approximation that the ratio of the production cross section is given by the ratio of partial widths even though  the QCD correction factors can differ in both cases. For example, the signal strength  is thus given by
\begin{equation}
\mu^{ggF}_{\gamma\gamma} = c_{hgg} c_{h\gamma\gamma} \frac{\Gamma_{SM}}{\Gamma_{NP}}
\end{equation}
The reduced couplings as well as all Higgs partial widths  are  computed within \micro~ and are provided in a SLHA file, 
\verb|HB.slha|.
The total width for the SM, necessary to compute Higgs branching ratios, can be computed from reduced couplings and the total width in the NP model.

 To obtain the  limits on the Higgs sector for models with more than one Higgs boson, 
the results for Higgs signal strengths in different channels described above can be
compared  to the latest results of the LHC. Alternatively an interface to the public code
HiggsBounds~\cite{Bechtle:2013gu} which allows to take into account 
the exclusion limits  provided by the experimental collaborations for many different channels is provided.
In \micro, the interface to HiggsBounds is realized via  SLHA files, using
the effective coupling option for specifying the input.  More specifically the file \verb|HB.slha| contains  
the normalized  couplings squared of the Higgs to all SM particles,
including the normalized couplings squared to  $\gamma\gamma,\gamma Z, gg$ ~\footnote{
The coupling $H_i \rightarrow \gamma Z$ will be included in a future release} as well all Higgs partial widths and the  top decay width.
A  reference value is defined for the couplings that do not exist in the standard model,
for example the reference value for the Z couplings to two Higgses,  $g_{H'HZ}^2= (e/2s_wc_w)^2$.
In the MSSM, analytical formulas for the  effective couplings normalized to the SM ones are used for $h,H,A\rightarrow  f\bar{f},WW,ZZ$ while  loop-induced couplings, $h,H,A \rightarrow \gamma\gamma, gg$, are computed using the formulas given above. The same procedure is used for most of the models we provide.
In the CPVMSSM , the only difference is that loop-induced couplings are computed with CPSuperH. 
In the NMSSM, we have two options for the Higgs sector :  collider constraints can be computed with NMSSMTools or with HiggsBounds. As for other models,   the normalized  Higgs couplings squared are provided in the \verb|Hb.slha| file. 

The SLHA file is then fed to HiggsBounds and 
the output of HiggsBounds is read using the slhaVal function~\cite{TASI}. Note that the output of HiggsBounds consist of two numbers, \verb|HBresult| and \verb|obsratio| which indicate whether a point has been excluded at the $95\%C.L.$ by one of the experimental results considered. 

\begin{tabular}{lll}
HBresult&obsratio&\\
0& $>1$ & parameter point is excluded\\
1& $<1$ & parameter point is not excluded\\
-1& $<0$ & invalid parameter point\\
\end{tabular}

\noindent
Information on the most constraining channel is also given as output. In Sectoon \ref{ExtPackages} we describe simple  tools for reading the  SLHA output of HiggsBounds.

In a generic model implemented by the user, the effective Lagrangian for $h\gamma\gamma$ and $hgg$ have to be included in the model file as described in section~\ref{sec:loop}. Furthermore if one wants to use the HiggsBounds interface, an SLHA file containing the normalized couplings squared must also be provided by the user. In models available in ~\micro, the following function can be used for this purpose

\noindent$\bullet$ \verb| HBblocks("filename") | \\
writes at the end of an SLHA file, the new Blocks that contain the effective couplings of the Higgses normalized to the SM ones.

\section{Dark matter models}

Several models are included in the \micro~ distribution:  the MSSM and some of its extensions
(NMSSM and CPVMSSM),  the little Higgs model (LHM)~\cite{Belyaev:2006jh},  a model with a right-handed neutrino (RHNM)~\cite{Belanger:2007dx},
the inert doublet model (IDM), a model with an extended scalar sector and a $Z_3$ symmetry(Z3ID)~\cite{Belanger:2012vp,Belanger:2012zr} as well as a model with a fourth generation of leptons (SM4). 
All but the last three models were available in previous versions. All models have been extended to include effective $h_igg$ and $h_i\gamma\gamma$ vertices for all Higgses in the model and to output a file containing reduced Higgs couplings ({\tt HB.slha}) - the only exception is the RHNM which features SM Higgs couplings.  Both the MSSM and NMSSM have been updated, we briefly describe the main new features of these models. A more complete description of all models will appear as a separate publication~\cite{micromodels}. A model independent calculation of the DM signals is also available.

\subsection{MSSM}

The model files were regenerated using SLHAplus~\cite{Belanger:2010st} and feature an improved Higgs sector which includes
\begin{itemize}
\item{} Effective vertices for $Hgg$ and $H\gamma\gamma$ where $H=h,H,A$.
\item{} Threshold corrections for $\tau$'s in addition to those for b-quarks~\cite{Carena:1994bv,Pierce:1996zz}.
\item{} Improved Higgs effective potential:  as in previous versions we write an effective Lagrangian to introduce in a gauge invariant manner  the higher-order corrections to the Higgs masses and couplings. The effective scalar potential contains seven parameters ($\lambda_i$)~\cite{Boudjema:2001ii}, 
analytic formulas are used to compute $\lambda_1,\lambda_3,\lambda_6$~\cite{Carena:1995wu,Djouadi:1997yw} other $\lambda$'s are computed from these and  the four Higgs masses, $m_h,m_H,m_A,m_{H^+}$ given by the spectrum calculator. This improvement has an impact on the Higgs  self-couplings, the effect on the DM relic density is however negligible.
\item{} The mass limit function checks that the Higgs mass is in the range $123<m_h<128$~GeV consistent with the LHC results and allowing for a theoretical uncertainty. A warning is issued if this is not the case. The Higgs couplings are not explicitly checked.
\end{itemize}
Other new features of the MSSM model
\begin{itemize}
\item{} The default spectrum generator has been updated to Suspect$\_2.4.1$
\item{} In addition to the CMSSM (called SUGRA), AMSB, and EWSB, the user can choose to work  with non-universal Higgs masses at the GUT scale (SUGRANUH). The scalar Higgs masses are not used as input parameters, rather they are replaced  by $\mu$, defined at the EW symmetry breaking scale and the pseudoscalar pole mass, $m_A$.  The input parameters are therefore  $m_0,m_{1/2},A_0,\tan\beta,\mu,m_A$ in addition to the optional standard model parameters $m_t,m_b(m_b),\alpha_s(m_Z)$. Defining input parameters at different scales  follows the standard in  SLHA2 ~\cite{Allanach:2008qq}.
\item{} We have updated the routine for $b\rightarrow s\gamma$ in order to reproduce the SM value at NNLO, default values for all parameters are given in the online manual.
\item{} We have added three flavor observables in the  Kaon and D-meson systems, the leptonic decays, 
$K^+\rightarrow \mu\nu$ and $D_s\rightarrow \mu\nu,\tau\nu$ as described below.
 Note that these constraints are also available via the SuperIso interface~\cite{Mahmoudi:2008tp}. 
\end{itemize}

The two-body  kaon leptonic decays receive a  contribution from the charged Higgs at tree-level. However  uncertainties in the Kaon parameters are very large, we therefore compute only a quantity that has much smaller uncertainties. This is defined as ~\cite{Antonelli:2008jg}
\begin{equation}
R_{l23}= \left[ 1-\frac{m_K^2}{m^2_{H^\pm}} \left( 1-\frac{m_d}{m_s}\right)  \frac{\tan\beta^2}{1+\epsilon_0\tan\beta}\right]
\end{equation}
with $m_K=0.494$~GeV and $m_d/m_s=1/20.2$.
Here the dependence on the supersymmetric parameters arises at higher order in perturbation theory (through the $\epsilon_0$ term) but can be large because of the $\tan\beta$ enhancement,
\begin{equation}
\epsilon_0=-\frac{2\alpha_s \mu}{3\pi \mglu} H_2(\frac{m^2_{{\tilde q}_L}}{\mglu^2},\frac{m^2_{{\tilde d}_R}}{\mglu^2})
\label{eq:epsilon}
\end{equation}
The  measured value for this  quantity is $R_{l23}= 1.004\pm 0.007$~\cite{Antonelli:2008jg}. 

The branching ratios $D_s\rightarrow \tau\nu$ and $D_s\rightarrow \mu\nu$ also receive a contribution from the charged Higgs,~\cite{Akeroyd:2009tn}
\begin{eqnarray}
Br(D_s\rightarrow l \nu)=\frac{G_F^2}{8\pi} &|V_{cs}|^2 f^2_{D_s} m_l^2 m_{D_s} \tau_{D_s} \left( 1-\frac{m_l^2}{m_{D_s}^2}\right)^2 \times\nonumber\\
& \left[ 1+\frac{1}{1+\frac{m_s}{m_c}}\frac{m^2_{D_s} }{m^2_{H^\pm}} \left(1-\frac{m_s}{m_c}\frac{ \tan^2\beta}{1+\epsilon_0 \tan\beta}  \right)  \right]^2
\end{eqnarray}
with $\epsilon_0$ as in Eq.~\ref{eq:epsilon} with $m_{{\tilde d}_R}\rightarrow m_{{\tilde c}_R}$
These branching ratios were measured recently, ~\cite{Amhis:2012bh} 
\begin{eqnarray}
Br(D_s\rightarrow \tau \nu) &=&(5.44\pm 0.22)\times 10^{-2}\nonumber\\
Br(D_s\rightarrow \mu \nu)&=& (5.54\pm 0.24)\times 10^{-3}
\end{eqnarray}

\subsection{NMSSM}

The NMSSM model files have also been updated and a new interface to
NMSSMTools\_4.0~\cite{Ellwanger:2005dv} is provided.
Thus all new features of  NMSSMTools~\cite{Ellwanger:2005dv,Ellwanger:2006rn} are included, in particular the computation of reduced Higgs couplings. 
The main new features of \micro~ are:

\begin{itemize}
\item{}Effective vertices for $Hgg$ and $H\gamma\gamma$ where $H=h_i,a_j$, $i=1,3; j=1,2$.
\item{} Improved Higgs effective potential:  as in previous versions we write an effective Lagrangian to introduce in a gauge invariant manner  the higher-order corrections to the Higgs masses and couplings~\cite{Belanger:2005kh}. The effective scalar potential has been extended to contain more than 20 operators, $\lambda_i$'s~\cite{greg_these}\footnote{See also
Ref.~\cite{Chalons:2012qe} for a general parametrization of the effective scalar potential}. 
Analytic formulas are used to compute the $\lambda_i$'s which are then fed to \micro.
The new Higgs self-coupling can in some cases have a large  effect on the DM relic density.
\end{itemize}

\subsection{Model independent signals}

A model independent calculation of the DM observables is also available.
After specifying the DM mass, the cross sections for DM  spin dependent and  spin independent scattering on proton and neutron, the DM annihilation cross section times velocity at rest and the relative contribution of  each annihilation channel to the total DM annihilation cross section, one can compute the direct detection rate on   various nuclei, the fluxes for photons, neutrinos and antimatter resulting from DM annihilation in the galaxy and the neutrino/muonfluxes in neutrino telescopes.  

The functions \verb|nuclearRecoilAux| for the computation of the event rate for direct detection    and
\verb|basicSpectra| which provides the $\gamma$, $p^-$, $e^+$ and $\nu$ spectra for different annihilation channels  are described in ~\cite{Belanger:2008sj}, \cite{ Belanger:2010gh}. The fluxes for photons, neutrinos, positrons and antiprotons from DM annihilation in the galaxy are computed using the same functions as in specific models. 
The new functions that allow  for  a model independent calculation of the neutrino  telescope  signal from
DM captured are 

\noindent $\bullet$ \verb|captureAux(f,forSun,csIp,csIn,csDp,csDn)|\\
calculates the number of DM particles captured per second assuming the cross sections
for  spin-independent and spin-dependent 
interactions with protons and neutrons   {\tt csIp, csIn, csDp, csDn} are
given as input parameters (in {\tt [pb]}). 
A negative value for one of the cross sections  is interpreted as a destructive
interference between the  proton and neutron amplitudes. The first two
parameters have the same meaning as in the {\tt neutrinoFlux} routine Section \ref{neutrino_functions}. The
result  depends implicitly on the global parameters {\tt rhoDM} and {\tt
Mcdm} in Table \ref{paramTab}.

\noindent
$\bullet$ \verb|basicNuSpectra(forSun,pdgN, outN,nu)|\\
calculates the $\nu_{\mu}$ and $\bar{\nu}_{\mu}$  spectra corresponding to  DM annihilating into particles specified by the PDG code {\tt pdgN}. Effects of interaction with Sun/Earth medium as well as neutrino oscillation  are taken into account \cite{Cirelli:2005gh}.
{\tt outN} should be chosen 1 for muon neutrino and -1 for anti-neutrino. The resulting spectrum is stored in the array 
  \verb|nu| with NZ=250 elements  which can be checked by the {\tt SpectdNdE(E,nu)}
function.

The annihilation rate for this
model independent analysis  is   obtained from the capture rate  using Eq.~\ref{eq:capture}.  
The  function{\tt basicNuSpectra} described above  can be used to  obtain the neutrino fluxes
assuming  some  user defined  value for each channel branching ratio   in Eq.(\ref{eq:nu_spectra}).

The files {\verb|main.c/F|} in the directory {\tt mdlIndep}  contain an example of the calculation of the direct detection,  indirect detection
and neutrino telescope signals using the routines described in this section. 
The numerical input data in this sample file corresponds to 'MSSM/mssmh.dat'.

\section{External codes in \micro~}
\label{ExtPackages}

The subdirectory {\tt Packages}   in \micro$\_3.1$  contains external public codes such as 
{\tt SuSpect\cite{Djouadi:2002ze}, CPsuperH\cite{Lee:2012wa}, HiggsBounds-4.0.0\cite{Bechtle:2013gu},  LoopTools-2.1\cite{Hahn:1998yk}}.   
These external codes are not compiled  during the  general installation of \micro, rather they are compiled 
either when installing a model requiring one of these codes 
 or in runtime of \micro~ executables. The  ~\micro~ interface to these codes (except for LoopTools) is done via files written following the SLHA convention.
Routines of the SLHAplus package~\cite{Belanger:2010st} included in \micro~ can be used to read the information contained in these files, as in previous versions the following function is available\\

\noindent
$\bullet$ \verb|slhaRead(filename,mode)|\\
reads all or  part of the data  from the file \verb|filename|.
\verb|mode| is an integer which determines which part of the data should be read from the file, 
\verb|mode= 1*m1+2*m2+4*m4+8*m8+16*m16|  where
\begin{verbatim}
  m1 = 0/1 -   overwrites all/keeps old data    
  m2 = 0/1 -   ignore errors in input file/ stop in case of error                                                                            
  m4 = 0/1 -   read DECAY /do not read   DECAY
  m8 = 0/1 -   read BLOCK/do not  read   BLOCK
  m16 = 0/1 -  read QNUMBERS/do not  read   QNUMBERS
\end{verbatim}

For example \verb|mode=20| (\verb|m4=1,m16=1|) is an instruction to overwrite all previous data and 
read only the information stored in the BLOCK sections of
\verb|filename|. In the same manner \verb|mode=25=1+8+16| is an instruction to add information from DECAY 
to the data obtained previously.

 The function \verb|slhaRead| returns the values:
\begin{verbatim}
  0  - successful reading
 -1  - can not open file
 -2  - invalid data as indicated by SPINFO
 -3  - no data
 n>0 - wrong file format at line n
\end{verbatim}

In SLHA, information is stored in blocks identifiable with the convention:
a line starting with the  keyword "BLOCK"   followed by the  name of the  block. 
Each block then contains a list of records, each on a separate line started with space symbol, that give information on physical parameters.
The structure of a block record is not defined in SLHA. In the first SLHA implementations, the structure of a block
was the following  :  key parameters  defining the variables  followed by  some  numerical parameters  giving the  value of these variables.
For example,
\begin{verbatim}
BLOCK MASS
  6                     173.2  # top quark mass 
  1000022     1.34362460e+02   # ~neutralino(1) 
\end{verbatim}
New codes have actually introduced new formats for blocks, for example in the  {\it Flavour Les Houches Accord} \cite{Mahmoudi:2010iz}
key parameters are specified both before and after the numerical parameters,
\begin{verbatim}
Block FOBS # Flavour observables
# ParentPDG type     value        q   NDA ID1  ID2 ID3 ... comment 
   5          1    2.83249483e-04 0   2    3    22  # BR(b->s gamma)
\end{verbatim}

The  SLHAplus package~\cite{Belanger:2010st} was updated  to  read blocks with any format, in addition to the function \verb|slhaVal|, we have introduced the function \\
\noindent
$\bullet$\verb|  slhaValFormat(char * Block, double Q, char * format)|\\
which allows to extract  numerical information corresponding to the key parameter
described in format. The {\it format } parameter is  actually  a C-style format. 
For instance in the  "BLOCK MASS"  above, the mass of the
neutralino can be obtained by \\
\verb| MNE1=slhaValFormat("MASS", 0., " 1000022 %lf");|\\
while in the "BLOCK FOBS",  the  branching $b \to s\gamma$ can be obtained by\\
\verb|B2sg=slhaValFormat("FOBS", 0., "5 1 %lf  %*f 2 3 22");|\\
The parameter {\it Q} is meaningful only for blocks
which feature a scale dependence. 
Note that the number of space symbols in {\it format} is irrelevant  since it has 
the same properties as the C {\it  scanf} function.

A block can also contain a textual information. For example, in
{\tt HIGGSBOUNDS} a block  contains the following records, 

{\scriptsize
\begin{verbatim}
Block HiggsBoundsResults 
#CHANNELTYPE 1: channel with the highest statistical sensitivity
    1           1         328                  # channel id number
    1           2           1                  # HBresult
    1           3  0.72692779334500290         # obsratio
    1           4           1                  # ncombined
    1           5 ||(p p)->h+..., h=1 where h is SM-like (CMS-PAS-HIG-12-008)|| # text description of channel
\end{verbatim}
}

\noindent 
In particular, the last record contains  the  name of the channel which gives the strongest constraint on the Higgs.  To extract the name
of this channel one can use the new function\\

\noindent
$\bullet$ \verb!slhaSTRFormat("HiggsBoundsResults","1 5 || %[^|]||",channel);!\\
which will write the channel name  in text parameter {\it channel}.

The following packages have been  attached to \micro~ using the SLHAplus interface: SOFTSUSY~\cite{Allanach:2001kg}, SPHENO~\cite{Porod:2003um}, ISAJET~\cite{Paige:2003mg}, SuperIso~\cite{Mahmoudi:2008tp}, SUSY-HIT~\cite{Djouadi:2006bz}, HDECAY~\cite{Djouadi:1997yw}, SusyBSG~\cite{Degrassi:2007kj}.
 
Note that the {\tt make clean}  command launched in  the \micro~ root directory 
subsequently  calls {\tt make clean } for all sub-directories  in the 
 {\tt Packages} directory.

\section{Installation and running the code}
\subsection{Installation.}
To   download  \micro~, go to    \\  
\verb|     http://lapth.cnrs.fr/micromegas|\\
and unpack the file received, \verb|micromegas_|\VERSION\verb|.tgz|, with the command\\
\verb|     tar -xvzf micromegas_|\VERSION\verb|.tgz|\\
This should create the directory \verb|micromegas_|\VERSION\verb|/|. 
In case of problems and questions please send a mail to \verb|micromegas@lapth.cnrs.fr| .

To compile   \calchep~ and \micro~ go to the \micro~ directory
and launch\\
\verb|     gmake|\\
The command {\tt make} should work like {\tt gmake} if the latter is not available.
In principle  \micro~  defines automatically the names of {\it C} and {\it
Fortran} compilers and the flags for
compilation. If you meet a  problem, open the file which contain the compiler specifications, 
\verb|CalcHEP_src/FlagsForSh|,
 improve it, and launch {\tt [g]make} 
again. 

 After compilation of \micro~ one has to compile
the executable to compute DM related observables in a specific model. To
do this, go to the model directory, say MSSM,  and launch\\
\verb|     [g]make main=main.c|\\
It should generate the executable {\tt main}. In the same manner\\
\verb|     gmake main=|{\it filename}.{\it ext}\\ 
generates the executable {\tt filename}  based on the source file {\it
filename.ext}.
For {\it ext}  we support 3 options: {\it 'c'} , {\it 'F'}, {\it 'cpp'} which correspond to
{\tt C}, {\tt FORTRAN} and {\tt C++} sources.
{\tt [g]make} called  in the model directory automatically  launches {\tt [g]make}
in subdirectories {\it lib} and {\it work} to compile \\
 \verb|     lib/aLib.a|   - library of auxiliary model functions, e.g. constraints,\\
 \verb|     work/work_aux.a| - library of model particles, free and dependent parameters.\\

Default versions of {\it main.c/F}  programs need some arguments
which have to be specified in command lines. If launched without
arguments {\it main} explains which parameter are needed. 
As a rule  {\it main}  needs  the name of a file containing the
numerical values of the free parameters of the model. 

To add a new model to \micro~ one has to launch the command \\
\verb|  ./newProject |{\it modelName} \\
which  creates the directory {\it modelName} which contains all subdirectories and
commands required. Then  one has to prepare the CalcHEP model files and place them  in the sub-directory 
{\tt work/models/}, furthermore auxiliary codes used in
the model have to be included  in the sub-directory {\tt lib/}.

\subsection{Module structure of main program}
Each model included in \micro~  is accompanied with sample files for
C and Fortran programs which call \micro~ routines, the {\it main.c}, {\it main.F}
files\footnote{The C++ version of the sample main program can be obtained by copying {\it
main.c} to {\it main.cpp}} .  
These files   consist of
several modules enclosed between the instructions
\begin{verbatim}
#ifdef KEYWORD
  ....................
#endif
\end{verbatim}
where {\tt KEYWORD} is the name of the module, all 
modules are completely independent. The user can comment or
uncomment any set of {\it define} instructions to suit his/her need.
The following modules are available for all models: 

\begin{verbatim}
MASSES_INFO   //Masses and widths of Higgs particles and masses
of particles in the odd sector.
CONSTRAINTS   //Checks various constraints.     
HIGGSBOUNDS   //Calls HiggsBounds to constrain the Higgs sector.
OMEGA         //DM Relic density.          
INDIRECT_DETECTION //Signals of DM annihilation in the galactic halo.   
RESET_FORMFACTORS  //Redefinition of nucleon Form Factors. 
CDM_NUCLEON   //Amplitudes and cross sections for DM-nucleon scattering. 
CDM_NUCLEUS   //Cross section and recoil energy distribution for
DM-nucleus scattering. 
NEUTRINO      //Neutrino flux from DM captured in the Sun/Earth
and the corresponding muon flux.     
DECAYS        //Decay widths and branching ratios.   
CROSS_SECTIONS//Cross sections for $2\to2$ processes.  
CLEAN         //Cleans intermediate files.
\end{verbatim}

For the three  MSSM-like models,   a module to calculate neutralino annihilation in $\gamma \gamma$ and $\gamma Z$ is also available, it is driven by the 
\verb|#define LoopGAMMA| instruction.

All models also contain the preprocessor  command  \verb|#define SHOWPLOTS|  which generates various plots in \verb|CalcHEP|. By default it is switched off. 

A complete description of all  \micro~ functions is available in the online manual, see also ~\cite{TASI}. This updated manual is also provided with the code, the file \verb|manual3.pdf| can be  found  in the 
 \micro\_3.X/man directory.

\subsection{Sample output}
\label{examples}

Running the main.c sample file in the \micro/MSSM directory choosing the options 
\verb|MASSES_INFO,CONSTRAINTS,OMEGA|, \verb|INDIRECT_DETECTION|, \verb|CDM_NUCLEON,NEUTRINO|,\verb|DECAYS| and with the mssmh.par input file will 
lead to the following output.

\begin{verbatim}
========= EWSB scale input =========
 Spectrum calculator is suspect
Initial file  "mssmh.par"
The following parapeters keep default values:
  alfSMZ=1.1720E-01      MW=8.0200E+01      MZ=9.1300E+01      Ml=1.7770E+00
    McMc=1.2700E+00    MbMb=4.2300E+00     Mtp=1.7350E+02      Au=0.0000E+00
      Ad=0.0000E+00      wt=1.4420E+00      wZ=2.5020E+00      wW=2.0940E+00
 SU_read_leshouches: end of file
 OUTPUT in SLHA format in suspect2_lha.out 
 RUN TERMINATED : OUTPUT in suspect2.out
Warnings from spectrum calculator:
 .....none

Dark matter candidate is '~o1' with spin=1/2  mass=1.89E+02

~o1 = 0.950*bino -0.062*wino +0.257*higgsino1 -0.167*higgsino2

=== MASSES OF HIGGS AND SUSY PARTICLES: ===
Higgs masses and widths
      h   126.08 3.49E-03
      H   699.89 5.16E+00
     H3   700.00 5.80E+00
     H+   704.62 4.74E+00

Masses of odd sector Particles:
~o1  : MNE1  =   189.0 || ~l1  : MSl1  =   198.6 || ~eR  : MSeR  =   204.8 
~mR  : MSmR  =   204.8 || ~1+  : MC1   =   283.8 || ~o2  : MNE2  =   290.0 
~o3  : MNE3  =   310.3 || ~2+  : MC2   =   444.0 || ~o4  : MNE4  =   444.2 
~ne  : MSne  =   496.0 || ~nm  : MSnm  =   496.0 || ~nl  : MSnl  =   496.0 
~eL  : MSeL  =   502.0 || ~mL  : MSmL  =   502.0 || ~l2  : MSl2  =   502.5 
~t1  : MSt1  =  1367.6 || ~b1  : MSb1  =  1551.8 || ~b2  : MSb2  =  1565.7 
~t2  : MSt2  =  1655.7 || ~g   : MSG   =  1900.2 || ~uL  : MSuL  =  2045.0 
~cL  : MScL  =  2045.0 || ~uR  : MSuR  =  2045.3 || ~cR  : MScR  =  2045.3 
~dR  : MSdR  =  2045.8 || ~sR  : MSsR  =  2045.8 || ~dL  : MSdL  =  2046.5 
~sL  : MSsL  =  2046.5 || 


==== Physical Constraints: =====
deltartho=3.56E-04
gmuon=1.15E-09
bsgnlo=3.05E-04 ( SM 3.25E-04 )
bsmumu=3.38E-09
btaunu=9.56E-01
dtaunu=5.17E-02  dmunu=5.33E-03
Rl23=9.998166E-01
MassLimits OK

==== Calculation of relic density =====
Xf=2.65e+01 Omega=1.03e-01

Channels which contribute to 1/(omega) more than 1%.
Relative contributions in % are displayed
 18% ~o1 ~o1 ->t T 
 16% ~o1 ~l1 ->A l 
 10% ~o1 ~o1 ->l L 
  7% ~l1 ~l1 ->l l 
  5% ~o1 ~l1 ->Z l 
  5% ~o1 ~eR ->A e 
  5% ~o1 ~mR ->A m 
  4% ~o1 ~o1 ->b B 
  4% ~o1 ~o1 ->e E 
  4% ~o1 ~o1 ->m M 
  3% ~eR ~l1 ->e l 
  3% ~mR ~l1 ->m l 
  2% ~o1 ~o1 ->W+ W- 
  2% ~l1 ~L1 ->A A 
  1% ~o1 ~eR ->Z e 
  1% ~o1 ~mR ->Z m 
  1% ~l1 ~L1 ->A Z 
  1% ~o1 ~o1 ->Z Z 

==== Indirect detection =======
    Channel          vcs[cm^3/s]
==================================
 annihilation cross section 9.22E-27 cm^3/s
  ~o1,~o1 -> t T        5.34E-27
  ~o1,~o1 -> l L        1.90E-27
  ~o1,~o1 -> b B        1.06E-27
  ~o1,~o1 -> W+ W-      5.57E-28
  ~o1,~o1 -> Z Z        2.56E-28
  ~o1,~o1 -> Z h        1.10E-28
sigmav=9.22E-27[cm^3/s]
Photon flux  for angle of sight f=0.00[rad]
and spherical region described by cone with angle 0.0349[rad]
Photon flux = 1.61E-14[cm^2 s GeV]^{-1} for E=94.5[GeV]
Positron flux  =  5.61E-13[cm^2 sr s GeV]^{-1} for E=94.5[GeV] 
Antiproton flux  =  6.34E-13[cm^2 sr s GeV]^{-1} for E=94.5[GeV] 

==== Calculation of CDM-nucleons amplitudes  =====
CDM-nucleon micrOMEGAs amplitudes:
proton:  SI  -3.547E-09  SD  -2.115E-07
neutron: SI  -3.634E-09  SD  1.862E-07
CDM-nucleon cross sections[pb]:
 proton  SI 5.446E-09  SD 5.807E-05
 neutron SI 5.715E-09  SD 4.503E-05

===============Neutrino Telescope=======  for  Sun
 E>1.0E+00 GeV neutrino flux       5.783577E+10 [1/Year/km^2] 
 E>1.0E+00 GeV anti-neutrino flux  6.224851E+10 [1/Year/km^2]
 E>1.0E+00 GeV Upward muon flux    2.258430E+02 [1/Year/km^2]
 E>1.0E+00 GeV Contained muon flux 1.217184E+03 [1/Year/km^3]

================= Decays ==============

h :   total width=4.564910E-03 
 and Branchings:
2.549952E-02 h -> Z,Z
2.098720E-01 h -> W+,W-
2.147891E-03 h -> A,A
6.913403E-02 h -> G,G
6.332254E-02 h -> l,L
1.367769E-04 h -> u,U
1.509459E-04 h -> d,D
2.669341E-02 h -> c,C
1.509459E-04 h -> s,S
6.028919E-01 h -> b,B

~o2 :   total width=1.357505E-01 
 and Branchings:
4.451636E-02 ~o2 -> Z,~o1
1.225739E-01 ~o2 -> E,~eR
1.225739E-01 ~o2 -> e,~ER
1.225739E-01 ~o2 -> M,~mR
1.225739E-01 ~o2 -> m,~MR
2.325939E-01 ~o2 -> L,~l1
2.325939E-01 ~o2 -> l,~L1
\end{verbatim}

\section*{Acknowledgements}
We thank C. Hugonie for providing the new NMSSMTools interface. We also thank A. Anandakrishnan, J. Da Silva,  G. Drieu La Rochelle, B. Dumont,  A. Goudelis, and S. Kulkarni for testing the beta versions of the code.
This work was  supported in part by the LIA-TCAP of CNRS,  by the French ANR, {\it Programme BLANC}/DMAstro-LHC and was performed within the Labex ENIGMASS.
The work of AP  was  also supported by the Russian foundation for Basic Research, grant
RFBR-12-02-93108-CNRSL-a.
GB and AP also thank the Galileo Galilei Institute for Theoretical Physics (GGI Florence) for hospitality and
the INFN for partial support.

\appendix
\section*{Appendix}

\section{Analytical solution of neutrino flux}

When DM is not self-conjugate, Eqs.~\ref{eq:ndot:asym} for the number density can be solved analytically
if we assume that  equilibrium is reached and that the evaporation rate is negligible.

Define
\begin{equation}
\beta= \frac{C_\chi}{C_{\bar\chi}}  \;\;\; ; \alpha=\frac{A_{\chi\chi}}{A_{\chi\bar\chi}}  \;\;\; x=\frac{N_\chi}{N_{\bar\chi}}
\end{equation}
Then after equilibrium is reached the annihilation rate is determined by the capture rates as well as $\alpha$, here we assume $ A_{\bar\chi\bar\chi} = A_{\chi\chi}$,

\begin{eqnarray}
\Gamma_{\chi\bar\chi} & =&\frac{C_{\chi}}{1+\alpha x}\\\nonumber
\Gamma_{\chi\chi} & =&\frac{C_\chi}{2}\left(\frac{ x \alpha }{1+\alpha x}\right)= C_{\bar\chi} \frac{x}{x+\alpha}\\\nonumber
\Gamma_{\bar\chi\bar\chi} & =&\frac{1}{2}C_{\bar\chi}\frac{\alpha }{1+\alpha x}  
\end{eqnarray}

where 
\begin{equation}
x=\frac{1}{2\alpha} \left[\beta-1 \pm \left((\beta-1)^2+4\alpha^2\beta\right)^{1/2}\right]
\end{equation}

\section{ Global Parameters}
\label{sec:global_parameters}

 A list of the
global parameters  and their default values  are given  in Tables
\ref{paramTab}, \ref{FFTab}. 
The numerical value for any of these parameters can be simply reset anywhere in the code. 

\begin{table}[htbp]
 \caption{Global parameters of \micro}
 \label{paramTab}
\begin{center}
\begin{tabular}{|l|l|l|l|l|}
\hline
  Name      &default value & units &  comments \\  \hline
  Mcdm      &              &  GeV  & Mass of Dark Matter particle           \\ 
  deltaY     &  0          &           & Difference of DM/anti-DM abundances \\
  DMasymm    &  0          &           & Asymmetry between relic density of  DM/anti-DM\\ 
K\_dif      & 0.0112     & kpc$^2$/Myr & The normalized diffusion coefficient\\
L\_dif      & 4           & kpc       & Vertical size of Galaxy diffisive halo \\
Delta\_dif   & 0.7        &           &Slope of diffision coefficient\\ 
Tau\_dif    & $10^{16}$   &   s       &Electron energy loss time\\
Vc\_dif     & 0           &  km/s     &  Convective Galactic vind \\
Fermi\_a    &  0.52        &  fm   & nuclei  surface thickness \\
Fermi\_b    &  -0.6        &  fm   &  parameters to set the nuclei radius with  \\    
Fermi\_c    &  1.23        &  fm   &  $R_A=c A^{1/3} +b$ \\ 
Rsun        & 8.5          & kpc   & Distance from Sun to center of Galaxy\\
Rdisk       & 20           & kpc   & Radius of galactic diffusion disk \\
rhoDM       &  0.3         & GeV/$cm^3$ & Dark Matter density at Rsun\\
Vearth      &  225.2   & km/s     & Galaxy velocity of Earth     \\
\hline
\end{tabular}
\end{center}
\end{table}

\begin{table}[]
 \caption{Global parameters of \micro :  nucleon quark form factors}
 \label{FFTab}
\begin{center}
\begin{tabular}{|l|l|l|l|l|l|}
\hline
 \multicolumn{2}{|c|}{Proton}& \multicolumn{2}{|c|}{Neutron} & \\ \hline
  Name      &  value       &  Name      &  value     &  comments \\  \hline
ScalarFFPd  &  0.0191     &ScalarFFNd  &  0.0273  & \\
ScalarFFPu  &  0.0153     &ScalarFFNu  &  0.011 & Scalar form factor \\
ScalarFFPs  &  0.0447      &ScalarFFNs  &  0.0447   & \\
\hline
pVectorFFPd &  -0.427      &pVectorFFNd & \phantom{-}0.842    & \\
pVectorFFPu &   \phantom{-}0.842      &pVectorFFNu &  -0.427   & Axial-vector form factor\\
pVectorFFPs &  -0.085      &pVectorFFNs &  -0.085   & \\
\hline
SigmaFFPd   &  -0.23       &SigmaFFNd   &  \phantom{-}0.84     & \\
SigmaFFPu   &   \phantom{-}0.84       &SigmaFFNu   &  -0.23    & Tensor form factor\\
SigmaFFPs   &   -0.046     &SigmaFFNs   &  -0.046   & \\
\hline
\end{tabular}
\end{center}
\end{table}

\newpage

\providecommand{\href}[2]{#2}\begingroup\raggedright\endgroup


\begin{thebibliography}{100}

\bibitem{Ade:2013lta}
{\bf Planck} Collaboration, P.~Ade {\em et al.}, ``{Planck 2013 results. XVI.
  Cosmological parameters},''
\href{http://www.arXiv.org/abs/1303.5076}{{\tt 1303.5076}}.

\bibitem{Bernabei:2010mq}
R.~Bernabei {\em et al.}, ``{New results from DAMA/LIBRA},'' {\em Eur. Phys.
  J.} {\bf C67} (2010) 39--49,
\href{http://www.arXiv.org/abs/1002.1028}{{\tt 1002.1028}}.

\bibitem{Agnese:2013rvf}
{\bf CDMS} Collaboration, R.~Agnese {\em et al.}, ``{Dark Matter Search Results
  Using the Silicon Detectors of CDMS II},'' {\em Phys.Rev.Lett.} (2013)
\href{http://www.arXiv.org/abs/1304.4279}{{\tt 1304.4279}}.

\bibitem{Aalseth:2011wp}
C.~E. Aalseth {\em et al.}, ``{Search for an Annual Modulation in a P-type
  Point Contact Germanium Dark Matter Detector},'' {\em Phys. Rev. Lett.} {\bf
  107} (2011) 141301,
\href{http://www.arXiv.org/abs/1106.0650}{{\tt 1106.0650}}.

\bibitem{Angloher:2011uu}
G.~Angloher, M.~Bauer, I.~Bavykina, A.~Bento, C.~Bucci, {\em et al.},
  ``{Results from 730 kg days of the CRESST-II Dark Matter Search},'' {\em
  Eur.Phys.J.} {\bf C72} (2012) 1971,
\href{http://www.arXiv.org/abs/1109.0702}{{\tt 1109.0702}}.

\bibitem{Aprile:2012nq}
{\bf XENON100} Collaboration, E.~Aprile {\em et al.}, ``{Dark Matter Results
  from 225 Live Days of XENON100 Data},''
\href{http://www.arXiv.org/abs/1207.5988}{{\tt 1207.5988}}.

\bibitem{Behnke:2008zza}
{\bf COUPP} Collaboration, E.~Behnke {\em et al.}, ``{Improved Spin-Dependent
  WIMP Limits from a Bubble Chamber},'' {\em Science} {\bf 319} (2008)
  933--936,
\href{http://www.arXiv.org/abs/0804.2886}{{\tt 0804.2886}}.

\bibitem{Archambault:2009sm}
S.~Archambault {\em et al.}, ``{Dark Matter Spin-Dependent Limits for WIMP
  Interactions on 19-F by PICASSO},'' {\em Phys. Lett.} {\bf B682} (2009)
  185--192,
\href{http://www.arXiv.org/abs/0907.0307}{{\tt 0907.0307}}.

\bibitem{Ackermann:2011wa}
{\bf Fermi-LAT} Collaboration, M.~Ackermann {\em et al.}, ``{Constraining Dark
  Matter Models from a Combined Analysis of Milky Way Satellites with the Fermi
  Large Area Telescope},'' {\em Phys.Rev.Lett.} {\bf 107} (2011) 241302,
\href{http://www.arXiv.org/abs/1108.3546}{{\tt 1108.3546}}.


\bibitem{Bringmann:2012vr}
  T.~Bringmann, X.~Huang, A.~Ibarra, S.~Vogl and C.~Weniger,
  JCAP {\bf 1207} (2012) 054
  [arXiv:1203.1312 [hep-ph]].
  
  \bibitem{Weniger:2012tx}
C.~Weniger, ``{A Tentative Gamma-Ray Line from Dark Matter Annihilation at the
  Fermi Large Area Telescope},'' {\em JCAP} {\bf 1208} (2012) 007,
\href{http://www.arXiv.org/abs/1204.2797}{{\tt 1204.2797}}.


\bibitem{FermiLAT:2011ab}
{\bf Fermi LAT} Collaboration, M.~Ackermann {\em et al.}, ``{Measurement of
  separate cosmic-ray electron and positron spectra with the Fermi Large Area
  Telescope},'' {\em Phys.Rev.Lett.} {\bf 108} (2012) 011103,
\href{http://www.arXiv.org/abs/1109.0521}{{\tt 1109.0521}}.

\bibitem{Adriani:2008zr}
{\bf PAMELA} Collaboration, O.~Adriani {\em et al.}, ``{An anomalous positron
  abundance in cosmic rays with energies 1.5-100 GeV},'' {\em Nature} {\bf 458}
  (2009) 607--609,
\href{http://www.arXiv.org/abs/0810.4995}{{\tt 0810.4995}}.

\bibitem{Aguilar:2013qda}
{\bf AMS} Collaboration, M.~Aguilar {\em et al.}, ``{First Result from the
  Alpha Magnetic Spectrometer on the International Space Station: Precision
  Measurement of the Positron Fraction in Primary Cosmic Rays of 0.5Ð-350
  GeV},'' {\em Phys.Rev.Lett.} {\bf 110} (2013), no.~14,
141102.

\bibitem{Adriani:2008zq}
O.~Adriani {\em et al.}, ``{A new measurement of the antiproton-to-proton flux
  ratio up to 100 GeV in the cosmic radiation},'' {\em Phys. Rev. Lett.} {\bf
  102} (2009) 051101,
\href{http://www.arXiv.org/abs/0810.4994}{{\tt 0810.4994}}.

\bibitem{Tanaka:2011uf}
{\bf Super-Kamiokande} Collaboration, T.~Tanaka {\em et al.}, ``{An Indirect
  Search for WIMPs in the Sun using 3109.6 days of upward-going muons in
  Super-Kamiokande},'' {\em Astrophys.J.} {\bf 742} (2011) 78,
\href{http://www.arXiv.org/abs/1108.3384}{{\tt 1108.3384}}.

\bibitem{Zornoza:2012xn}
J.~Zornoza, ``{Search for Dark Matter in the Sun with the ANTARES Neutrino
  Telescope in the CMSSM and mUED frameworks},''
\href{http://www.arXiv.org/abs/1204.5290}{{\tt 1204.5290}}.

\bibitem{IceCube:2011aj}
{\bf IceCube} Collaboration, R.~Abbasi {\em et al.}, ``{Multi-year search for
  dark matter annihilations in the Sun with the AMANDA-II and IceCube
  detectors},'' {\em Phys.Rev.} {\bf D85} (2012) 042002,
\href{http://www.arXiv.org/abs/1112.1840}{{\tt 1112.1840}}.

\bibitem{Aad:2012tfa}
{\bf ATLAS} Collaboration, G.~Aad {\em et al.}, ``{Observation of a new
  particle in the search for the Standard Model Higgs boson with the ATLAS
  detector at the LHC},'' {\em Phys.Lett.} {\bf B716} (2012) 1--29,
\href{http://www.arXiv.org/abs/1207.7214}{{\tt 1207.7214}}.

\bibitem{Chatrchyan:2012ufa}
{\bf CMS} Collaboration, S.~Chatrchyan {\em et al.}, ``{Observation of a new
  boson at a mass of 125 GeV with the CMS experiment at the LHC},'' {\em
  Phys.Lett.} {\bf B716} (2012) 30--61,
\href{http://www.arXiv.org/abs/1207.7235}{{\tt 1207.7235}}.

\bibitem{Aad:2012fw}
{\bf ATLAS} Collaboration, G.~Aad {\em et al.}, ``{Search for dark matter
  candidates and large extra dimensions in events with a photon and missing
  transverse momentum in $pp$ collision data at $\sqrt{s}=7$ TeV with the ATLAS
  detector},'' {\em Phys.Rev.Lett.} {\bf 110} (2013) 011802,
\href{http://www.arXiv.org/abs/1209.4625}{{\tt 1209.4625}}.

\bibitem{ATLAS:2012ky}
{\bf ATLAS} Collaboration, G.~Aad {\em et al.}, ``{Search for dark matter
  candidates and large extra dimensions in events with a jet and missing
  transverse momentum with the ATLAS detector},'' {\em JHEP} {\bf 1304} (2013)
  075,
\href{http://www.arXiv.org/abs/1210.4491}{{\tt 1210.4491}}.

\bibitem{Chatrchyan:2012me}
{\bf CMS Collaboration} Collaboration, S.~Chatrchyan {\em et al.}, ``{Search
  for dark matter and large extra dimensions in monojet events in $pp$
  collisions at $\sqrt{s}=7$ TeV},'' {\em JHEP} {\bf 1209} (2012) 094,
\href{http://www.arXiv.org/abs/1206.5663}{{\tt 1206.5663}}.

\bibitem{Gondolo:2004sc}
P.~Gondolo {\em et al.}, ``{DarkSUSY: Computing supersymmetric dark matter
  properties numerically},'' {\em JCAP} {\bf 0407} (2004) 008,
\href{http://www.arXiv.org/abs/astro-ph/0406204}{{\tt astro-ph/0406204}}.

\bibitem{Baer:2004qq}
H.~Baer, A.~Belyaev, T.~Krupovnickas, and J.~O'Farrill, ``{Indirect, direct and
  collider detection of neutralino dark matter},'' {\em JCAP} {\bf 0408} (2004)
  005,
\href{http://www.arXiv.org/abs/hep-ph/0405210}{{\tt hep-ph/0405210}}.

\bibitem{Arbey:2009gu}
A.~Arbey and F.~Mahmoudi, ``{SuperIso Relic: A Program for calculating relic
  density and flavor physics observables in Supersymmetry},'' {\em
  Comput.Phys.Commun.} {\bf 181} (2010) 1277--1292,
\href{http://www.arXiv.org/abs/0906.0369}{{\tt 0906.0369}}.

\bibitem{Semenov:2008jy}
A.~Semenov, ``{LanHEP - a package for the automatic generation of Feynman rules
  in field theory. Version 3.0},'' {\em Comput. Phys. Commun.} {\bf 180} (2009)
  431--454,
\href{http://www.arXiv.org/abs/0805.0555}{{\tt 0805.0555}}.

\bibitem{Barbieri:2006dq}
R.~Barbieri, L.~J. Hall, and V.~S. Rychkov, ``{Improved naturalness with a
  heavy Higgs: An Alternative road to LHC physics},'' {\em Phys.Rev.} {\bf D74}
  (2006) 015007,
\href{http://www.arXiv.org/abs/hep-ph/0603188}{{\tt hep-ph/0603188}}.

\bibitem{Belanger:2012vp}
G.~Belanger, K.~Kannike, A.~Pukhov, and M.~Raidal, ``{Impact of
  semi-annihilations on dark matter phenomenology - an example of $Z_N$
  symmetric scalar dark matter},'' {\em JCAP} {\bf 04} (2012) 010,
\href{http://www.arXiv.org/abs/1202.2962}{{\tt 1202.2962}}.

\bibitem{Ellwanger:2005dv}
U.~Ellwanger and C.~Hugonie, ``{NMHDECAY 2.0: An Updated program for sparticle
  masses, Higgs masses, couplings and decay widths in the NMSSM},'' {\em
  Comput.Phys.Commun.} {\bf 175} (2006) 290--303,
\href{http://www.arXiv.org/abs/hep-ph/0508022}{{\tt hep-ph/0508022}}.

\bibitem{Ellwanger:2006rn}
U.~Ellwanger and C.~Hugonie, ``{NMSPEC: A Fortran code for the sparticle and
  Higgs masses in the NMSSM with GUT scale boundary conditions},'' {\em
  Comput.Phys.Commun.} {\bf 177} (2007) 399--407,
\href{http://www.arXiv.org/abs/hep-ph/0612134}{{\tt hep-ph/0612134}}.

\bibitem{Djouadi:2002ze}
A.~Djouadi, J.-L. Kneur, and G.~Moultaka, ``{SuSpect: A Fortran code for the
  supersymmetric and Higgs particle spectrum in the MSSM},'' {\em Comput. Phys.
  Commun.} {\bf 176} (2007) 426--455,
\href{http://www.arXiv.org/abs/hep-ph/0211331}{{\tt hep-ph/0211331}}.

\bibitem{Lee:2012wa}
J.~Lee, M.~Carena, J.~Ellis, A.~Pilaftsis, and C.~Wagner, ``{CPsuperH2.3: an
  Updated Tool for Phenomenology in the MSSM with Explicit CP Violation},''
  {\em Comput.Phys.Commun.} {\bf 184} (2013) 1220--1233,
\href{http://www.arXiv.org/abs/1208.2212}{{\tt 1208.2212}}.

\bibitem{TASI}
G.~B\'elanger, F.~Boudjema, and A.~Pukhov, {\em micrOMEGAs: A Package for
  Calculation of Dark Matter Properties in Generic Model of Particle
  Interaction}, ch.~12, pp.~739--790.
\newblock World Scientific, 2012.
\newblock
  \href{http://www.arXiv.org/abs/http://www.worldscientific.com/doi/pdf/10.1142/9789814390163\_0012}{{\tt
  http://www.worldscientific.com/doi/pdf/10.1142/9789814390163\_0012}}.

\bibitem{Belanger:2007dx}
G.~Belanger, A.~Pukhov, and G.~Servant, ``{Dirac Neutrino Dark Matter},'' {\em
  JCAP} {\bf 0801} (2008) 009,
\href{http://www.arXiv.org/abs/0706.0526}{{\tt 0706.0526}}.

\bibitem{Barger:2008jx}
V.~Barger, P.~Langacker, M.~McCaskey, M.~Ramsey-Musolf, and G.~Shaughnessy,
  ``{Complex Singlet Extension of the Standard Model},'' {\em Phys.Rev.} {\bf
  D79} (2009) 015018,
\href{http://www.arXiv.org/abs/0811.0393}{{\tt 0811.0393}}.

\bibitem{Belanger:2012zr}
G.~Belanger, K.~Kannike, A.~Pukhov, and M.~Raidal, ``{$Z_3$ Scalar Singlet Dark
  Matter},''
\href{http://www.arXiv.org/abs/1211.1014}{{\tt 1211.1014}}.

\bibitem{Graesser:2011wi}
M.~L. Graesser, I.~M. Shoemaker, and L.~Vecchi, ``{Asymmetric WIMP dark
  matter},'' {\em JHEP} {\bf 1110} (2011) 110,
\href{http://www.arXiv.org/abs/1103.2771}{{\tt 1103.2771}}.

\bibitem{Iminniyaz:2011yp}
H.~Iminniyaz, M.~Drees, and X.~Chen, ``{Relic Abundance of Asymmetric Dark
  Matter},'' {\em JCAP} {\bf 1107} (2011) 003,
\href{http://www.arXiv.org/abs/1104.5548}{{\tt 1104.5548}}.

\bibitem{Ellwanger:2012yg}
U.~Ellwanger and P.~Mitropoulos, ``{Upper Bounds on Asymmetric Dark Matter Self
  Annihilation Cross Sections},'' {\em JCAP} {\bf 1207} (2012) 024,
\href{http://www.arXiv.org/abs/1205.0673}{{\tt 1205.0673}}.

\bibitem{Belanger:2004yn}
G.~Belanger, F.~Boudjema, A.~Pukhov, and A.~Semenov, ``{micrOMEGAs: Version
  1.3},'' {\em Comput. Phys. Commun.} {\bf 174} (2006) 577--604,
\href{http://www.arXiv.org/abs/hep-ph/0405253}{{\tt hep-ph/0405253}}.

\bibitem{Belanger:2013kya}
G.~Belanger, B.~Dumont, U.~Ellwanger, J.~Gunion, and S.~Kraml, ``{Status of
  invisible Higgs decays},''
\href{http://www.arXiv.org/abs/1302.5694}{{\tt 1302.5694}}.

\bibitem{Giardino:2013bma}
P.~P. Giardino, K.~Kannike, I.~Masina, M.~Raidal, and A.~Strumia, ``{The
  universal Higgs fit},''
\href{http://www.arXiv.org/abs/1303.3570}{{\tt 1303.3570}}.

\bibitem{Hambye:2008bq}
T.~Hambye, ``{Hidden vector dark matter},'' {\em JHEP} {\bf 0901} (2009) 028,
\href{http://www.arXiv.org/abs/0811.0172}{{\tt 0811.0172}}.

\bibitem{D'Eramo:2010ep}
F.~D'Eramo and J.~Thaler, ``{Semi-annihilation of Dark Matter},'' {\em JHEP}
  {\bf 1006} (2010) 109,
\href{http://www.arXiv.org/abs/1003.5912}{{\tt 1003.5912}}.

\bibitem{Hindmarsh:2005ix}
M.~Hindmarsh and O.~Philipsen, ``{WIMP dark matter and the QCD equation of
  state},'' {\em Phys.Rev.} {\bf D71} (2005) 087302,
\href{http://www.arXiv.org/abs/hep-ph/0501232}{{\tt hep-ph/0501232}}.

\bibitem{Olive:1980wz}
K.~A. Olive, D.~N. Schramm, and G.~Steigman, ``{Limits on New Superweakly
  Interacting Particles from Primordial Nucleosynthesis},'' {\em Nucl. Phys.}
  {\bf B180} (1981)
497.

\bibitem{Srednicki:1988ce}
M.~Srednicki, R.~Watkins, and K.~A. Olive, ``{Calculations of relic densities
  in the early universe},'' {\em Nucl. Phys.} {\bf B310} (1988)
693.

\bibitem{Hinshaw:2012fq}
G.~Hinshaw, D.~Larson, E.~Komatsu, D.~Spergel, C.~Bennett, {\em et al.},
  ``{Nine-Year Wilkinson Microwave Anisotropy Probe (WMAP) Observations:
  Cosmological Parameter Results},''
\href{http://www.arXiv.org/abs/1212.5226}{{\tt 1212.5226}}.

\bibitem{Yaguna:2010hn}
C.~E. Yaguna, ``{Large contributions to dark matter annihilation from
  three-body final states},'' {\em Phys.Rev.} {\bf D81} (2010) 075024,
\href{http://www.arXiv.org/abs/1003.2730}{{\tt 1003.2730}}.

\bibitem{Goudelis:2013uca}
A.~Goudelis, B.~Herrmann, and O.~StŒl, ``{Dark matter in the Inert Doublet
  Model after the discovery of a Higgs-like boson at the LHC},''
\href{http://www.arXiv.org/abs/1303.3010}{{\tt 1303.3010}}.

\bibitem{Honorez:2010re}
L.~Lopez~Honorez and C.~E. Yaguna, ``{The inert doublet model of dark matter
  revisited},'' {\em JHEP} {\bf 09} (2010) 046,
\href{http://www.arXiv.org/abs/1003.3125}{{\tt 1003.3125}}.

\bibitem{Belanger:2006is}
G.~Belanger, F.~Boudjema, A.~Pukhov, and A.~Semenov, ``{micrOMEGAs2.0: A
  program to calculate the relic density of dark matter in a generic model},''
  {\em Comput. Phys. Commun.} {\bf 176} (2007) 367--382,
\href{http://www.arXiv.org/abs/hep-ph/0607059}{{\tt hep-ph/0607059}}.

\bibitem{Belanger:2001fz}
G.~Belanger, F.~Boudjema, A.~Pukhov, and A.~Semenov, ``{micrOMEGAs: A program
  for calculating the relic density in the MSSM},'' {\em Comput. Phys. Commun.}
  {\bf 149} (2002) 103--120,
\href{http://www.arXiv.org/abs/hep-ph/0112278}{{\tt hep-ph/0112278}}.

\bibitem{Abbasi:2012ws}
{\bf IceCube} Collaboration, R.~Abbasi {\em et al.}, ``{Search for Neutrinos
  from Annihilating Dark Matter in the Direction of the Galactic Center with
  the 40-String IceCube Neutrino Observatory},''
\href{http://www.arXiv.org/abs/1210.3557}{{\tt 1210.3557}}.

\bibitem{Boliev:2013ai}
M.~Boliev, S.~Demidov, S.~Mikheyev, and O.~Suvorova, ``{Search for muon signal
  from dark matter annihilations in the Sun with the Baksan Underground
  Scintillator Telescope for 24.12 years},''
\href{http://www.arXiv.org/abs/1301.1138}{{\tt 1301.1138}}.

\bibitem{Ha:2012np}
{\bf IceCube} Collaboration, C.~H. Ha, ``{The First Year IceCube-DeepCore
  Results},'' {\em J.Phys.Conf.Ser.} {\bf 375} (2012) 052034,
\href{http://www.arXiv.org/abs/1201.0801}{{\tt 1201.0801}}.

\bibitem{Gould:1987ir}
A.~Gould, ``{Resonant Enhancements in WIMP Capture by the Earth},'' {\em
  Astrophys.J.} {\bf 321} (1987)
571.

\bibitem{Asplund:2004eu}
M.~Asplund, N.~Grevesse, and J.~Sauval, ``{The Solar chemical composition},''
  {\em Nucl.Phys.} {\bf A777} (2006) 1--4,
\href{http://www.arXiv.org/abs/astro-ph/0410214}{{\tt astro-ph/0410214}}.

\bibitem{Geochemistry}
W.~McDonough, {\em Compositional Model for the Earth's Core}, vol.~2,
  p.~547–568.
\newblock Elsevier Ltd, 2003.

\bibitem{Griest:1986yu}
K.~Griest and D.~Seckel, ``{Cosmic Asymmetry, Neutrinos and the Sun},'' {\em
  Nucl. Phys.} {\bf B283} (1987)
681.

\bibitem{Gould:1987ju}
A.~Gould, ``{WIMP distribution in and evaporation from the Sun},'' {\em
  Astrophys.J.} {\bf 321} (1987)
560.

\bibitem{Gould:1989tu}
A.~Gould, ``{Evaporation of WIMPs with arbitrary cross-sections},'' {\em
  Astrophys. J.}
(1989).

\bibitem{Cirelli:2005gh}
M.~Cirelli {\em et al.}, ``{Spectra of neutrinos from dark matter
  annihilations},'' {\em Nucl. Phys.} {\bf B727} (2005) 99--138,
\href{http://www.arXiv.org/abs/hep-ph/0506298}{{\tt hep-ph/0506298}}.

\bibitem{Belanger:2010cd}
G.~Belanger, M.~Kakizaki, E.~K. Park, S.~Kraml, and A.~Pukhov, ``{Light mixed
  sneutrinos as thermal dark matter},'' {\em JCAP} {\bf 1011} (2010) 017,
\href{http://www.arXiv.org/abs/1008.0580}{{\tt 1008.0580}}.

\bibitem{Erkoca:2009by}
A.~E. Erkoca, M.~H. Reno, and I.~Sarcevic, ``{Muon Fluxes From Dark Matter
  Annihilation},'' {\em Phys. Rev.} {\bf D80} (2009) 043514,
\href{http://www.arXiv.org/abs/0906.4364}{{\tt 0906.4364}}.

\bibitem{Belanger:2008sj}
G.~Belanger, F.~Boudjema, A.~Pukhov, and A.~Semenov, ``{Dark matter direct
  detection rate in a generic model with micrOMEGAs2.1},'' {\em Comput. Phys.
  Commun.} {\bf 180} (2009) 747--767,
\href{http://www.arXiv.org/abs/0803.2360}{{\tt 0803.2360}}.

\bibitem{Belanger:2010gh}
G.~Belanger {\em et al.}, ``{Indirect search for dark matter with
  micrOMEGAs2.4},'' {\em Comput. Phys. Commun.} {\bf 182} (2011) 842--856,
\href{http://www.arXiv.org/abs/1004.1092}{{\tt 1004.1092}}.

\bibitem{Boudjema:2005hb}
F.~Boudjema, A.~Semenov, and D.~Temes, ``{Self-annihilation of the neutralino
  dark matter into two photons or a Z and a photon in the MSSM},'' {\em
  Phys.Rev.} {\bf D72} (2005) 055024,
\href{http://www.arXiv.org/abs/hep-ph/0507127}{{\tt hep-ph/0507127}}.

\bibitem{Baro:2008bg}
N.~Baro, F.~Boudjema, and A.~Semenov, ``{Automatised full one-loop
  renormalisation of the MSSM. I. The Higgs sector, the issue of tan(beta) and
  gauge invariance},'' {\em Phys.Rev.} {\bf D78} (2008) 115003,
\href{http://www.arXiv.org/abs/0807.4668}{{\tt 0807.4668}}.

\bibitem{Baro:2009gn}
N.~Baro and F.~Boudjema, ``{Automatised full one-loop renormalisation of the
  MSSM II: The chargino-neutralino sector, the sfermion sector and some
  applications},'' {\em Phys.Rev.} {\bf D80} (2009) 076010,
\href{http://www.arXiv.org/abs/0906.1665}{{\tt 0906.1665}}.

\bibitem{Chalons:2011ia}
G.~Chalons and A.~Semenov, ``{Loop-induced photon spectral lines from
  neutralino annihilation in the NMSSM},'' {\em JHEP} {\bf 1112} (2011) 055,
\href{http://www.arXiv.org/abs/1110.2064}{{\tt 1110.2064}}.

\bibitem{Zhao:1995cp}
H.~Zhao, ``{Analytical Models For Galactic Nuclei},'' {\em Mon. Not. Roy.
  Astron. Soc.} {\bf 278} (1996) 488--496,
\href{http://www.arXiv.org/abs/astro-ph/9509122}{{\tt astro-ph/9509122}}.

\bibitem{Lavalle:2006vb}
J.~Lavalle, J.~Pochon, P.~Salati, and R.~Taillet, ``{Clumpiness of Dark Matter
  and Positron Annihilation Signal: Computing the odds of the Galactic
  Lottery},'' {\em Astron. Astrophys.} {\bf 462} (2007) 827--848,
\href{http://www.arXiv.org/abs/astro-ph/0603796}{{\tt astro-ph/0603796}}.

\bibitem{Bottino:1999ei}
A.~Bottino, F.~Donato, N.~Fornengo, and S.~Scopel, ``{Implications for relic
  neutralinos of the theoretical uncertainties in the neutralino nucleon
  cross-section},'' {\em Astropart.Phys.} {\bf 13} (2000) 215--225,
\href{http://www.arXiv.org/abs/hep-ph/9909228}{{\tt hep-ph/9909228}}.

\bibitem{Ellis:2008hf}
J.~R. Ellis, K.~A. Olive, and C.~Savage, ``{Hadronic Uncertainties in the
  Elastic Scattering of Supersymmetric Dark Matter},'' {\em Phys.Rev.} {\bf
  D77} (2008) 065026,
\href{http://www.arXiv.org/abs/0801.3656}{{\tt 0801.3656}}.

\bibitem{Thomas:2012tg}
A.~Thomas, P.~Shanahan, and R.~Young, ``{Strangeness in the nucleon: what have
  we learned?},'' {\em Nuovo Cim.} {\bf C035N04} (2012) 3--10,
\href{http://www.arXiv.org/abs/1202.6407}{{\tt 1202.6407}}.

\bibitem{Beringer:1900zz}
{\bf Particle Data Group} Collaboration, J.~Beringer {\em et al.}, ``{Review of
  Particle Physics (RPP)},'' {\em Phys.Rev.} {\bf D86} (2012)
010001.

\bibitem{Oksuzian:2012rzb}
{\bf JLQCD} Collaboration, H.~Ohki {\em et al.}, ``{Nucleon strange quark
  content from $N_f=2+1$ lattice QCD with exact chiral symmetry},''
\href{http://www.arXiv.org/abs/1208.4185}{{\tt 1208.4185}}.

\bibitem{Engelhardt:2012gd}
M.~Engelhardt, ``{Strange quark contributions to nucleon mass and spin from
  lattice QCD},'' {\em Phys.Rev.} {\bf D86} (2012) 114510,
\href{http://www.arXiv.org/abs/1210.0025}{{\tt 1210.0025}}.

\bibitem{Freeman:2012ry}
{\bf MILC} Collaboration, W.~Freeman and D.~Toussaint, ``{The intrinsic
  strangeness and charm of the nucleon using improved staggered fermions},''
\href{http://www.arXiv.org/abs/1204.3866}{{\tt 1204.3866}}.

\bibitem{Young:2009zb}
R.~Young and A.~Thomas, ``{Octet baryon masses and sigma terms from an SU(3)
  chiral extrapolation},'' {\em Phys.Rev.} {\bf D81} (2010) 014503,
\href{http://www.arXiv.org/abs/0901.3310}{{\tt 0901.3310}}.

\bibitem{Durr:2011mp}
S.~Durr, Z.~Fodor, T.~Hemmert, C.~Hoelbling, J.~Frison, {\em et al.}, ``{Sigma
  term and strangeness content of octet baryons},'' {\em Phys.Rev.} {\bf D85}
  (2012) 014509,
\href{http://www.arXiv.org/abs/1109.4265}{{\tt 1109.4265}}.

\bibitem{Horsley:2011wr}
R.~Horsley, Y.~Nakamura, H.~Perlt, D.~Pleiter, P.~Rakow, {\em et al.},
  ``{Hyperon sigma terms for 2+1 quark flavours},'' {\em Phys.Rev.} {\bf D85}
  (2012) 034506,
\href{http://www.arXiv.org/abs/1110.4971}{{\tt 1110.4971}}.

\bibitem{Semke:2012gs}
A.~Semke and M.~Lutz, ``{Strangeness in the baryon ground states},'' {\em
  Phys.Lett.} {\bf B717} (2012) 242--247,
\href{http://www.arXiv.org/abs/1202.3556}{{\tt 1202.3556}}.

\bibitem{Shanahan:2012wh}
P.~Shanahan, A.~Thomas, and R.~Young, ``{Sigma terms from an SU(3) chiral
  extrapolation},''
\href{http://www.arXiv.org/abs/1205.5365}{{\tt 1205.5365}}.

\bibitem{Ren:2012aj}
X.-L. Ren, L.~Geng, J.~M. Camalich, J.~Meng, and H.~Toki, ``{Octet baryon
  masses in next-to-next-to-next-to-leading order covariant baryon chiral
  perturbation theory},''
\href{http://www.arXiv.org/abs/1209.3641}{{\tt 1209.3641}}.

\bibitem{Junnarkar:2013ac}
P.~Junnarkar and A.~Walker-Loud, ``{The Scalar Strange Content of the Nucleon
  from Lattice QCD},''
\href{http://www.arXiv.org/abs/1301.1114}{{\tt 1301.1114}}.

\bibitem{Bali:2012qs}
G.~Bali, P.~Bruns, S.~Collins, M.~Deka, B.~Glasle, {\em et al.}, ``{Nucleon
  mass and sigma term from lattice QCD with two light fermion flavors},'' {\em
  Nucl.Phys.} {\bf B866} (2013) 1--25,
\href{http://www.arXiv.org/abs/1206.7034}{{\tt 1206.7034}}.

\bibitem{Belyaev:2012qa}
A.~Belyaev, N.~D. Christensen, and A.~Pukhov, ``{CalcHEP 3.4 for collider
  physics within and beyond the Standard Model},''
\href{http://www.arXiv.org/abs/1207.6082}{{\tt 1207.6082}}.

\bibitem{Djouadi:2005gi}
A.~Djouadi, ``{The Anatomy of electro-weak symmetry breaking. I: The Higgs
  boson in the standard model},'' {\em Phys.Rept.} {\bf 457} (2008) 1--216,
\href{http://www.arXiv.org/abs/hep-ph/0503172}{{\tt hep-ph/0503172}}.

\bibitem{Djouadi:2005gj}
A.~Djouadi, ``{The Anatomy of electro-weak symmetry breaking. II. The Higgs
  bosons in the minimal supersymmetric model},'' {\em Phys.Rept.} {\bf 459}
  (2008) 1--241,
\href{http://www.arXiv.org/abs/hep-ph/0503173}{{\tt hep-ph/0503173}}.

\bibitem{Djouadi:1997yw}
A.~Djouadi, J.~Kalinowski, and M.~Spira, ``{HDECAY: A Program for Higgs boson
  decays in the standard model and its supersymmetric extension},'' {\em
  Comput.Phys.Commun.} {\bf 108} (1998) 56--74,
\href{http://www.arXiv.org/abs/hep-ph/9704448}{{\tt hep-ph/9704448}}.

\bibitem{Baikov:2006ch}
P.~Baikov and K.~Chetyrkin, ``{Higgs Decay into Hadrons to Order
  alpha**5(s)},'' {\em Phys.Rev.Lett.} {\bf 97} (2006) 061803,
\href{http://www.arXiv.org/abs/hep-ph/0604194}{{\tt hep-ph/0604194}}.

\bibitem{Bechtle:2011sb}
P.~Bechtle, O.~Brein, S.~Heinemeyer, G.~Weiglein, and K.~E. Williams,
  ``{HiggsBounds 2.0.0: Confronting Neutral and Charged Higgs Sector
  Predictions with Exclusion Bounds from LEP and the Tevatron},'' {\em
  Comput.Phys.Commun.} {\bf 182} (2011) 2605--2631,
\href{http://www.arXiv.org/abs/1102.1898}{{\tt 1102.1898}}.

\bibitem{micromodels}
G.~B. {\it et al}, ``Implementaion of new physics models in ~\micro.'' in
  preparation.

\bibitem{Belanger:2010st}
G.~Belanger, N.~D. Christensen, A.~Pukhov, and A.~Semenov, ``{SLHAplus: a
  library for implementing extensions of the standard model},'' {\em
  Comput.Phys.Commun.} {\bf 182} (2011) 763--774,
\href{http://www.arXiv.org/abs/1008.0181}{{\tt 1008.0181}}.

\bibitem{Carena:1994bv}
M.~S. Carena, M.~Olechowski, S.~Pokorski, and C.~Wagner, ``{Electroweak
  symmetry breaking and bottom - top Yukawa unification},'' {\em Nucl.Phys.}
  {\bf B426} (1994) 269--300,
\href{http://www.arXiv.org/abs/hep-ph/9402253}{{\tt hep-ph/9402253}}.

\bibitem{Pierce:1996zz}
D.~M. Pierce, J.~A. Bagger, K.~T. Matchev, and R.-j. Zhang, ``{Precision
  corrections in the minimal supersymmetric standard model},'' {\em Nucl.
  Phys.} {\bf B491} (1997) 3--67,
\href{http://www.arXiv.org/abs/hep-ph/9606211}{{\tt hep-ph/9606211}}.

\bibitem{Boudjema:2001ii}
F.~Boudjema and A.~Semenov, ``{Measurements of the SUSY Higgs selfcouplings and
  the reconstruction of the Higgs potential},'' {\em Phys.Rev.} {\bf D66}
  (2002) 095007,
\href{http://www.arXiv.org/abs/hep-ph/0201219}{{\tt hep-ph/0201219}}.

\bibitem{Carena:1995wu}
M.~S. Carena, M.~Quiros, and C.~Wagner, ``{Effective potential methods and the
  Higgs mass spectrum in the MSSM},'' {\em Nucl.Phys.} {\bf B461} (1996)
  407--436,
\href{http://www.arXiv.org/abs/hep-ph/9508343}{{\tt hep-ph/9508343}}.

\bibitem{Allanach:2008qq}
B.~Allanach, C.~Balazs, G.~Belanger, M.~Bernhardt, F.~Boudjema, {\em et al.},
  ``{SUSY Les Houches Accord 2},'' {\em Comput.Phys.Commun.} {\bf 180} (2009)
  8--25,
\href{http://www.arXiv.org/abs/0801.0045}{{\tt 0801.0045}}.

\bibitem{Mahmoudi:2008tp}
F.~Mahmoudi, ``{SuperIso v2.3: A Program for calculating flavor physics
  observables in Supersymmetry},'' {\em Comput.Phys.Commun.} {\bf 180} (2009)
  1579--1613,
\href{http://www.arXiv.org/abs/0808.3144}{{\tt 0808.3144}}.

\bibitem{Antonelli:2008jg}
{\bf FlaviaNet Working Group on Kaon Decays} Collaboration, M.~Antonelli {\em
  et al.}, ``{Precision tests of the Standard Model with leptonic and
  semileptonic kaon decays},''
\href{http://www.arXiv.org/abs/0801.1817}{{\tt 0801.1817}}.

\bibitem{Akeroyd:2009tn}
A.~Akeroyd and F.~Mahmoudi, ``{Constraints on charged Higgs bosons from
  $D(s)^\pm \rightarrow \mu^\pm \nu$ and $D(s)^\pm \rightarrow \tau^\pm
  \nu$},'' {\em JHEP} {\bf 0904} (2009) 121,
  \href{http://www.arXiv.org/abs/0902.2393}{{\tt 0902.2393}}.

\bibitem{Amhis:2012bh}
{\bf Heavy Flavor Averaging Group} Collaboration, Y.~Amhis {\em et al.},
  ``{Averages of B-Hadron, C-Hadron, and tau-lepton properties as of early
  2012},''
\href{http://www.arXiv.org/abs/1207.1158}{{\tt 1207.1158}}.

\bibitem{Belanger:2005kh}
G.~Belanger, F.~Boudjema, C.~Hugonie, A.~Pukhov, and A.~Semenov, ``{Relic
  density of dark matter in the NMSSM},'' {\em JCAP} {\bf 0509} (2005) 001,
\href{http://www.arXiv.org/abs/hep-ph/0505142}{{\tt hep-ph/0505142}}.

\bibitem{Hahn:1998yk}
T.~Hahn and M.~Perez-Victoria, ``{Automatized one loop calculations in
  four-dimensions and D-dimensions},'' {\em Comput.Phys.Commun.} {\bf 118}
  (1999) 153--165,
\href{http://www.arXiv.org/abs/hep-ph/9807565}{{\tt hep-ph/9807565}}.

\bibitem{Mahmoudi:2010iz}
F.~Mahmoudi, S.~Heinemeyer, A.~Arbey, A.~Bharucha, T.~Goto, {\em et al.},
  ``{Flavour Les Houches Accord: Interfacing Flavour related Codes},'' {\em
  Comput.Phys.Commun.} {\bf 183} (2012) 285--298,
\href{http://www.arXiv.org/abs/1008.0762}{{\tt 1008.0762}}.

\bibitem{Allanach:2001kg}
B.~C. Allanach, ``{SOFTSUSY: a program for calculating supersymmetric
  spectra},'' {\em Comput. Phys. Commun.} {\bf 143} (2002) 305--331,
\href{http://www.arXiv.org/abs/hep-ph/0104145}{{\tt hep-ph/0104145}}.

\bibitem{Porod:2003um}
W.~Porod, ``{SPheno, a program for calculating supersymmetric spectra, SUSY
  particle decays and SUSY particle production at e+ e- colliders},'' {\em
  Comput. Phys. Commun.} {\bf 153} (2003) 275--315,
\href{http://www.arXiv.org/abs/hep-ph/0301101}{{\tt hep-ph/0301101}}.

\bibitem{Paige:2003mg}
F.~E. Paige, S.~D. Protopopescu, H.~Baer, and X.~Tata, ``{ISAJET 7.69: A Monte
  Carlo event generator for pp, anti-p p, and e+e- reactions},''
\href{http://www.arXiv.org/abs/hep-ph/0312045}{{\tt hep-ph/0312045}}.

\bibitem{Djouadi:2006bz}
A.~Djouadi, M.~Muhlleitner, and M.~Spira, ``{Decays of supersymmetric
  particles: The Program SUSY-HIT (SUspect-SdecaY-Hdecay-InTerface)},'' {\em
  Acta Phys.Polon.} {\bf B38} (2007) 635--644,
\href{http://www.arXiv.org/abs/hep-ph/0609292}{{\tt hep-ph/0609292}}.

\bibitem{Degrassi:2007kj}
G.~Degrassi, P.~Gambino, and P.~Slavich, ``{SusyBSG: A Fortran code for $BR(B
  \to X(s) \gamma$ in the MSSM with Minimal Flavor Violation},'' {\em
  Comput.Phys.Commun.} {\bf 179} (2008) 759--771,
\href{http://www.arXiv.org/abs/0712.3265}{{\tt 0712.3265}}.

\end{thebibliography}
\end{document}